\documentclass[a4paper,11pt]{article}
\usepackage{fullpage,citesort}
\usepackage{amsmath,amssymb}
\usepackage{epsfig}

\newcommand{\tr}{\operatorname{tr}}
\newcommand{\Tr}{\operatorname{Tr}}
\newcommand{\Arctan}{\operatorname{Arctan}}

\newcommand{\diag}{\operatorname{diag}}
\newcommand{\negcdot}{\negmedspace\cdot\negthinspace}
\newcommand{\be}{\begin{equation}}
\newcommand{\ee}{\end{equation}}
\newcommand{\sls}{\not\negmedspace}

\begin{document}
\begin{titlepage}
\vspace{1cm}
\begin{center}
{\Large \bf  The Radiative  Leptonic Decays $D^0 \to  e^+ e^-\gamma,\mu^+\mu^-\gamma $ in the Standard Model and Beyond }\\

\vspace{1cm}
{\large \bf S. Fajfer$^{a,b}$,  P. Singer$^{c}$, J. Zupan$^a$\\}

\vspace{1cm}
{\it a) J. Stefan Institute, Jamova 39, P. O. Box 3000, 1001 Ljubljana,
Slovenia}
\vspace{.5cm}

{\it b)
Department of Physics, University of Ljubljana, Jadranska 19, 1000
Ljubljana,
Slovenia}
\vspace{.5cm}

{\it c) Department of Physics, Technion - Israel Institute  of
Technology,
Haifa 32000, Israel}

\end{center}

\vspace{0.25cm}

\centerline{\large \bf ABSTRACT}

\vspace{0.2cm}
We present a calculation of the rare decay modes $ D^0 \to e^+ e^- \gamma $ and $ D^0 \to \mu^+\mu^- \gamma  $  in the framework of Standard Model. For the short distance  part, we have derived QCD
corrections to the Wilson coefficients involved, including $C_9$. The latter
is found to be strongly suppressed by the corrections, leading to diminished values for the $c\to u l^+l^-$ branching ratios in the $10^{-10}$ range. Within SM the exclusive decays are dominated by long distance  effects. Nonresonant contributions are
estimated using heavy quark and chiral symmetries to be at the level of
$10\%$, compared to the contributions arising from $D^0\to V \gamma\to l^+l^-\gamma$, with $V=\rho, \omega, \phi$. The total SM branching ratio is predicted
to be in the range $(1-3)\times 10^{-9}$.  We also consider contributions coming from  MSSM with and without $R$ parity conservation. Effects from MSSM are significant only for the $R$-parity violating case. Such contributions enhance the branching ratio $D^0\to \mu^+\mu^-\gamma$ to $\lesssim 0.5 \times 10^{-7}$, based on appropriately allowed values for $C_9$ and $C_{10}$. This selects  $D^0\to \mu^+\mu^-\gamma$ as a possible  probe of new physics.  


\end{titlepage}

\section{Introduction}
The charm physics is entering an exciting era. The high statistics and an excellent quality of data at FOCUS experiment now allow, among others, for high precision studies of  charm semileptonic decays \cite{Link:2002ev}, determination of $D^{0,\pm}$ decay times below $1\%$ error level \cite{Link:2002bx}, as well as for searches of CP violation and rare $D$ decays \cite{Link:2000aw,Link:2000kr}. There is a very rich potential for  charm physics at B-factories, with both Belle and BaBar having an active program in charm studies \cite{Yabsley:2002,Williams:2002}. For instance, more than 120 milion charm pairs have already been produced at BaBar. This corresponds to more than 220 000 $D^*$-tagged $D^0$ decays, which will allow for precision lifetime and $D^0$ mixing analyses as well as for searches of rare charm decays \cite{Williams:2002}. An exciting charm physics program is under way also at CLEO, that  was recently able to  measure $\Gamma(D^*)$ for the first time  \cite{Ahmed:2001xc,Anastassov:2001cw}. Among the rare D
decays, the decays $D\to V \gamma$ and $D \to V(P) l^+ l^-$  are subjects of CLEO and FERMILAB searches \cite{E791,Johns:2002hd}. In the following years a great phenomenological impact  is expected from proposed CLEO-c physics programme. Next year more than 6 million tagged $D$ decays are expected to be measured. This will allow for precision charm branching ratio measurements and consequently improved measurements of CKM matrix elements also in $b$-sector, as well as for extensive studies of $D$-mixing, CP violation and rare decays in the charm sector \cite{Pedlar}.

Parallel to the experimental studies, there has been an ongoing theoretical effort to understand charm physics. A number of studies has focused on the rare charm decays \cite{Burdman:1995te,Bajc:1994ui,Greub:1996wn,Lebed:1999kq,Fajfer:1998dv,Fajfer:1998rz,Fajfer:2001ad,Fajfer:2002bq,Geng:2000if} and a possible impact of new physics on the predicted branching ratios  \cite{Bigi:1989hw,Burdman:2001tf,Fajfer:1999dq,Fajfer:2000zx,Fajfer:2001sa,Prelovsek:2000xy}. Note however, that in rare $D$ decays the nonperturbative physics of light quarks is expected to dominate the decay rates. Consider for instance the case of $c\to u\gamma$ transition that occurs only at  one loop level in the Standard Model. The contributions coming from $b,s,d$ quarks running in the loop are
\begin{equation}
{\cal M}(c\to u)=\sum_{q=d,s,b}V_{uq}^* V_{cq} {\cal M}_q \sim 
\left\{
\begin{array}{l@{\quad:\quad}l}
{\cal O}(\lambda^5 m_b^2)& b-\text{quark},\\
{\cal O}(\lambda m_s^2)& s-\text{quark},\\
{\cal O}(\lambda \Lambda_{QCD}^2)& d-\text{quark},
\end{array}
\right.
\end{equation}
where we have tentatively set $\Lambda_{QCD}$ instead of $m_u$ for the $u-$quark contribution, anticipating the size of nonperturbative effects. The situation is quite different from the $s\to d$ FCNC (e.g. $s\to d\nu \bar{\nu}$), where the same CKM hierarchy is present, but with the top quark replacing the $b$ quark. Since $b$ quark is much lighter than the top quark, it cannot surpass the $\lambda^4$ suppression. Thus the contributions from the heaviest, $b$-quark, are expected to be the least important. One can then expect that in rare $D$ decays the  nonperturbative long distance (LD) effects coming from the lighter two down quarks, $d,s$ will give the dominant contributions.

Since LD effects are difficult to control theoretically one would like to either find decay modes where LD effects are as small as possible and/or find observables where LD effects cancel. Such an observable was constructed in \cite{Fajfer:2000zx}, where $D^0\to (\rho, \omega) \gamma$ decays were considered. It was found that most of the LD effects cancel in the difference of the appropriately renormalized decay widths, making these channels a useful probe of new physics. Another interesting analysis is connected with decay modes $D\to (P, V) l^+l^-$, where $P=\pi, K, \eta$ are the light pseudoscalar mesons, $V=\rho, \omega, \dots$ are the light vector mesons and $l^+l^-$ is an electron or muon lepton pair. The decays have been estimated both in the SM and MSSM \cite{Fajfer:1998rz,Burdman:2001tf,Fajfer:2001sa,Schwartz:he,Singer:1996it}. In Ref. \cite{Burdman:2001tf} it was found that the experimental bounds on  $\text{Br}(D^+\to \pi^+ \mu^+\mu^-)$,  $\text{Br}(D^0\to \rho^0 \mu^+\mu^-)$ constrain the sizes of relevant trilinear $R$ parity violating coupling more stringently than analyses from other processes. Measurements of rare $D$ meson  decays can thus already now constrain new physics scenarios in the up-like quark sector.

In this paper we investigate the rare decays $D^0\to  e^+e^-\gamma$, $D^0\to  \mu^+\mu^-\gamma$ both in the Standard Model and in MSSM. A Standard Model  analysis of $D^0\to l^+l^- \gamma$ branching ratios neglecting QCD effects and LD transitions, has been made in Ref. \cite{Geng:2000if}, giving $\text{Br}(D^0\to l^+l^- \gamma)=6.3 \times  10^{-11}$.  However,   LD effects are expected to dominate the SM prediction similarly to the $D\to (P, V) l^+l^-$  decays. To evaluate the nonresonant LD effects we use the heavy quark effective theory combined with chiral perturbation theory (HQ$\chi$PT) \cite{Wise:hn}. This approach was used before for treating $D^*$ strong and
electromagnetic
decays \cite{Casalbuoni:1996pg,Guetta:1999vb}, as well as the leptonic and semileptonic decays of
$D$ meson (see \cite{Casalbuoni:1996pg} and references therein) and $D^0\to \gamma\gamma$ decay \cite{Fajfer:2001ad}. In addition, we include contributions of vector resonances in our analysis.

Another expectation based on the experience from $D\to (P, V) l^+l^-$  decays is that there are  possibly large contributions in  $D^0\to l^+l^- \gamma$ decays coming from SM extensions such as MSSM with R-parity violation. These expectations make  $D^0\to l^+l^- \gamma$ channels interesting from both experimental as well as from theoretical side.

Our calculations show that as a result of the LD contributions, these
decays would occur with a branching ratio of $(1-3)\times 10^{-9}$ in the SM, nearly
two orders of magnitude larger than in the previous estimate \cite{Geng:2000if}. Moreover,
MSSM with $R$-parity violation as presently restricted, allows for a
branching ratio of $D^0\to \mu^+\mu^-\gamma$ in the $10^{-7}$ range.

The paper is organized as follows. We start with the Standard Model prediction in section 2, where first  a discussion of renormalization group improved effective weak Lagrangian together with the calculation of $c\to ul^+l^-$ inclusive mode is given. This is then followed by the estimates of nonresonant as well as of resonant LD contributions to the decay width $D^0\to l^+l^- \gamma$. In section 3 we present possible effects of SUSY extensions of Standard Model. In appendices we collect some further details about the calculation of $c\to u$ effective weak Lagrangian, as well as the explicit formulae  of the calculations.

\section{Standard Model calculation}
\begin{figure}
\begin{center}
\epsfig{file=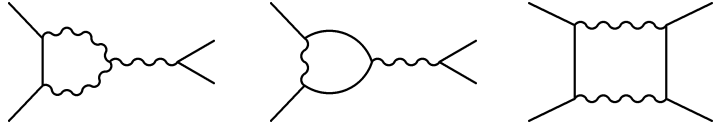, height=2.cm}
\caption{\footnotesize{The penguin and box diagrams contributing to $ D^0 \to l^+ l^-\gamma$ decay at quark level.}}\label{InamiLim}
\end{center}
\end{figure}

We will devote the first part of the present paper to the estimation of $ D^0 \to l^+ l^-\gamma$ decay width in the context of the Standard Model. At the quark level, this decay mode cannot proceed through tree diagrams and is thus induced only at the one loop level in the Standard Model. Possible quark diagrams  are shown on  Fig. \ref{InamiLim}. These then translate into an effective weak Lagrangian at the scale of $m_c$.
 
\subsection{Effective weak Lagrangian}\label{Eff-weak-lagr}
The effective Lagrangian describing the weak $c\to u$ transitions at the scale of $\mu=m_c$ is (see Appendix \ref{app-Wilson}, Eqs. \eqref{lagr-at-weak}-\eqref{lagr-at-bandc})
\begin{equation}
{\cal L}_{\text{eff}}=
-\frac{G_F}{\sqrt{2}} \big[ V_{cd}^* V_{ud} \sum_{i=1,2}C_i Q_i^d+ V_{cs}^* V_{us} \sum_{i=1,2}C_i Q_i^s-V_{cb}^* V_{ub} \sum_{i=3,\dots,10} C_iQ_i\big], \label{weak-eff-lagr}
\end{equation}
where
\begin{subequations}\label{list-oper}
\begin{align}
Q_1^q&=(\bar{u}^\alpha q^\beta)_{V-A} (\bar{q}^\beta c^\alpha)_{V-A} ,&  Q_2^q=&(\bar{u}q)_{V-A}(\bar{q}c)_{V-A},\\
Q_3&=(\bar{u}c)_{V-A}\sum_q (\bar{q}q)_{V-A}, & Q_4=&(\bar{u}^\alpha c^\beta)_{V-A}\sum_q (\bar{q}^\beta q^\alpha)_{V-A},\\
Q_5&=(\bar{u}c)_{V-A}\sum_q (\bar{q}q)_{V+A}, & Q_6=&(\bar{u}^\alpha c^\beta)_{V-A}\sum_q (\bar{q}^\beta q^\alpha)_{V+A},\\
Q_{7}&=\frac{e}{4\pi^2} m_c F_{\mu\nu} \bar{u} \sigma^{\mu\nu} P_R c, &Q_{8}=&\frac{g_s}{4\pi^2} m_c G^a_{\mu\nu} \bar{u} \sigma^{\mu\nu} T^a P_R c \label{Q7},\\
Q_9&=\frac{e^2}{16 \pi^2} (\bar{u}_L \gamma^\mu c_L) (\bar{l} \gamma_\mu l),
&Q_{10}=&\frac{e^2}{16 \pi^2} (\bar{u}_L \gamma^\mu c_L) (\bar{l} \gamma_\mu \gamma_5 l),\label{Q9}
\end{align}
\end{subequations}
with $q_L=P_L q$ and $P_{R,L}=(1\pm \gamma_5)/2$  the chirality projection operators, while we have suppressed the color indices in the currents of form $(\bar{q}q')=(\bar{q}^{\alpha}q'{}^\alpha)$. The sum over $q$ runs over the active quark-flavors. At scale $\mu\simeq m_c$ these are $q=u,d,s,c$. $C_i$ are Wilson coefficients to which QCD corrections are administered. We do not include in the analysis higher dimension operators.

Note, that the penguin operators $Q_{3,\dots,10}$ are proportional to $V_{cb}^* V_{ub}$ matrix elements in the effective weak Lagrangian \eqref{weak-eff-lagr}. In Wolfenstein parametrization this is $\sim \lambda^5$, which has to be compared to the CKM suppression of $Q_{1,2}$ operators, $V_{cs} V_{us}\sim \lambda$, where $\lambda=\sin{\theta_c}=0.22$. Penguin operators are thus greatly suppressed in $\Delta C=1$ transitions.  They are relevant only  in special observables such as $CP$ asymmetries \cite{Buccella:1994nf}. In the literature \cite{Geng:2000if,Burdman:2001tf,Fajfer:2001sa} as an estimate for $C_9(\mu_c)$ Wilson coefficient, the result   from  electroweak theory without QCD, $C_9^{\text{IL}}$ (where IL stands for Inami-Lim \cite{Inami:1980fz})  has been used. Since $C_9^{\text{IL}}$ is not $V_{ub}$ suppressed, it greatly overestimates the effect of $Q_9$ operator insertion on the predicted decay widths. We will thus devote the rest of this  section to clarify this point. 

The values of Wilson coefficients $C_1,\dots,C_{10}$ at scale $\mu=m_c$ are obtained by using the same method as in the existing calculations for $s\to d$ transitions \cite{Buchalla:1995vs,Buras:qa,Buras:xp} at leading (LO) and next-to-leading order (NLO). Application to $c\to u$ transition is straightforward, but some care  has to be taken when integrating out the $b$-quark at the intermediate step of renormalization group (RG) evolution. The charge of the intermediate $b$-quark is important for the matching  of electroweak $C_9$ Wilson coefficient. Since this calculation has not yet been performed we give  further details in appendix \ref{app-Wilson}. The Wilson coefficients $C_1,\dots, C_6$ for $c\to u$ transitions have been calculated already in \cite{Buccella:1994nf} at NLO, while the LO calculation of  $C_7$ has been presented in \cite{Burdman:1995te,Greub:1996wn}. The values of calculated Wilson coefficients are listed in Table \ref{tab-Wilson}. For a comparison the values of Wilson coefficients at LO order are given as well, but calculated with the two-loop evolution of the strong coupling constant. The values are given for the  central value of $\Lambda^{(5)}=216\pm25$ MeV and the matching scale $m_b=4.25$ GeV.  The one sigma change in $\Lambda^{(5)}$ corresponds to a change of about 10\% in $C_{1,\dots,6}$. We find a pronounced scale dependence for the $C_9$ coefficient below 1.5 GeV, as a consequence of large cancelations in the RG evolution equations. The situation is very similar to the case of coefficient $Z_{7V}$ in $K_L \to \pi^0 e^+e^-$ \cite{Buchalla:1995vs}. The LO value of $C_9$ even changes sign near $\mu \sim 1$ GeV, being positive for $\mu >1$ GeV. Note, that uncertainties in the value of the $C_9$ coefficient will not propagate into the decay rates as the $Q_9$ operator is $V_{ub}$ suppressed. Note also, that $Q_{10}$ does not mix with other operators due to chirality, so that $C_{10}(\mu_c)=C_{10}(\mu_W)\simeq 0$.

\begin{table} [h]
\begin{center}
\begin{tabular}{|l|c|c|c|c|c|c|c|c|} \hline

-&$\mu$(GeV) &  $C_1$& $C_2$ &  $C_3$ &  $C_4$ &  $C_5$ &  $C_6$  & $C_9$\\ \hline\hline
LO&$1.0$ &  $-0.64$& $1.34$ &  $0.016$ &  $-0.036$ &  $0.010$ &  $-0.046$  & $-0.07$\\ \hline
NLO&$1.0$ &  $-0.49$& $1.26$ &  $0.024$ &  $-0.060$ &  $0.015$ &  $-0.060$  & $-0.60$\\ \hline
NLO&$1.5$ &  $-0.37$& $1.18$ &  $0.013$ &  $-0.036$ &  $0.012$ &  $-0.033$ & $-0.13$\\ \hline
NLO&$2.0$ &  $-0.30$& $1.14$ &  $0.009$ &  $-0.025$ &  $0.009$ &  $-0.021$  & $-0.13$\\ \hline
 \end{tabular} 
 \caption{\footnotesize{Values of Wilson coefficients at scales $\mu=1.0 \; \text{GeV},1.5\; \text{GeV}, 2.0$ GeV, calculated at next-to-leading order (NLO) as explained in text. For a comparison in the first line the LO values are given at scale $\mu=1$ GeV, but calculated with two loop evolution of strong coupling constant. }}
\label{tab-Wilson}
\end{center}
\end{table}

As for the  $C_7$ Wilson coefficient, the leading order mixing of operators $Q_{7,8}$ with operators $Q_{1,\dots,6}$ vanishes.  It is only at two-loop that the anomalous dimension matrix has nonzero values mixing $C_{1,\dots,6}$ into $C_7$.  Since two-loop results are  scheme dependent, it is then customary to introduce an effective anomalous dimension matrix $\gamma^{(0)\text{eff}}$ \cite{Buras:xp}, which is scheme independent as is the case in leading order results. Using LO anomalous dimension matrix $\gamma^{(0)\text{eff}}$ and  NLO evolution for $\alpha_s$,   $m_b=4.25$ GeV, we arrive at (see also \cite{Greub:1996wn})
\begin{equation}
C_7^{\text{eff}}(1.0 \;\text{GeV})=0.13, \qquad C_7^{\text{eff}}(1.5\;\text{GeV})=0.087, \qquad C_7^{\text{eff}}(2.0\; \text{GeV})=0.065, \label{C7numbers}
\end{equation}

   Note that, as we already mentioned, the Wilson coefficient $C_9(\mu_c)$
has been estimated previously \cite{Geng:2000if,Burdman:2001tf,Fajfer:2001sa}  by using the result from
electroweak theory without QCD, i.e. taking  $C_9^{\text{IL}}$, based on the
(unproved) expectation that $C_9$ is not much affected by QCD corrections.
The leading order expression in terms of  $m_{d,s}^2/m_W^2$ is \footnote{For further details about the calculation see appendix \ref{app-Wilson}, where also a discussion regarding $C_{7,10}$ is presented}
\begin{equation}
C_9^{\text{IL}}\simeq -\lambda_s 16/9 \ln \big(m_s/m_d), \label{C9IL}
\end{equation}
where $\lambda_j=V_{cj}^* V_{uj}/(V_{cb}^* V_{ub})$. Using $m_s/m_d=17-22$ \cite{Hagiwara:pw}  we arrive at  the value $V_{cb}^*V_{ub} C_9^{\text{IL}}\simeq -V_{cs}^* V_{us} 16/9 \ln \big(m_s/m_d)=-1.13 \pm 0.06 $ which should be compared to  $V_{cb}^*V_{ub} C_9(\mu)\sim 10^{-4}$. The value of Wilson coefficient is thus four orders of magnitude smaller than the corresponding parameter obtained by neglecting QCD interactions! The reason for this discrepancy lies in the appearance of large logarithms $\ln(m_{d,s}/m_W)$ that  avoid the GIM suppression otherwise present in  $C_9$. It is exactly these large logarithms that RG evolution sums correctly. Since small scales of order  $m_{d,s}$  lie in the nonperturbative region of QCD, the approximation of using \eqref{C9IL} without QCD corrections is not
valid. 

The logarithm appearing in \eqref{C9IL} is exactly reproduced in the calculation of inclusive modes $c\to u l^+ l^-$,   if  mass-independent renormalization is used (see appendix C of \cite{Buras:1991jm}). To show this explicitly, we consider the calculation of $c\to u l^+l^-$ in the naive dimensional regularization (NDR).  The amplitude can be parametrized as 
\begin{equation}
M=- \frac{G_F}{\sqrt{2}}V_{cb}^* V_{ub}\left [ \hat{C}_7^{\text{eff}}\langle Q_7\rangle^{0}+\hat{C}_9^{\text{eff}}\langle Q_9\rangle^{0}+\hat{C}_{10}^{\text{eff}}\langle Q_{10}\rangle^{0}\right],
\end{equation}
with $\langle Q_{7,9,10}\rangle^0$ the tree level matrix elements of the operators. Note that $\hat{C}_{7,9,10}^{\text{eff}}$ are not Wilson coefficients but merely parametrize the invariant amplitude. The $\hat{C}_9^{\text{eff}}$ coefficient is dominated by the 1-loop contributions coming from  insertion of $Q_{1,2}^q$ operators, $q=d,s$.  The virtual photon is emitted from the intermediate $d,s$ quarks. This contribution is of order $\alpha_s^0$ and proportional to $V_{cq}^* V_{uq}$ and is thus only once Cabibbo suppressed.  Using existing results for $b\to s\l^+l^-$ at NLO \cite{Grinstein:1988me,Misiak:bc,Buras:1994dj},  we arrive at
\begin{equation}
V_{cb}^* V_{ub} \hat{C}_9^{\text{eff}}=2 V_{cs}^* V_{us} \left( h(z_s,\hat s)-h(z_d, \hat s)\right) \left( 3 C_1(m_c) +C_2(m_c)\right), \label{hatc9}
\end{equation}
with $z_q=m_q/m_c$, $\hat s=(m_{l^+l^-}/m_c)^2$ and $m_{l^+l^-}$ the  mass of the lepton pair, while
\begin{equation}
\begin{split}
h(z,s)=&-\frac{8}{9} \ln z+\frac{8}{27}+\frac{4}{9}x\\
&-\frac{2}{9}(2+x)\sqrt{|1-x|}
\left\{ 
\begin{aligned}
 \ln\left|\frac{\sqrt{1-x}+1}{\sqrt{1-x}-1}\right|-i \pi,\qquad &\text{for\;}x<1,\\
2 \Arctan\left(\frac{1}{\sqrt{x-1}}\right),\qquad &\text{for\;}x \ge 1,
\end{aligned}
\right.
\end{split}
\end{equation}
where $x=4z^2/s$.  In \eqref{hatc9} the contributions  suppressed by $V_{cb}^* V_{ub}$ are neglected. These include the tree-level contribution from $Q_9$ as well as 1-loop  contributions coming from insertions of QCD penguin operators $Q_{3,\dots 6}$. From expression  \eqref{hatc9} one should reproduce the Inami-Lim  result \eqref{C9IL}, when momenta and masses of external particles are set to zero. Taking the limit $m_{l^+l^-}\ll m_{d,s}$, one gets 
\begin{equation}
\lim_{\hat{s}\to 0}\left( h(z_s,\hat s)-h(z_d, \hat s)\right)\to -\frac{8}{9} \ln\left( \frac{m_s}{m_d}\right)
\end{equation}
Taking the values of $C_{1,2}$ Wilson coefficients at the weak scale $C_1\simeq 0$, $C_2 \simeq 1$ one arrives at the Inami-Lim result \eqref{C9IL}, as expected. Note, that the logarithm $\ln(m_d/m_s)$ in  \eqref{C9IL} arises from the insertion of $Q_{1,2}$ operators. Phenomenologically more interesting is the limit $m_{l^+l^-}\sim m_c\gg m_{d,s}$. In the limit $m_{l^+l^-}\to \infty$ the difference  $\left( h(z_s,\hat s)-h(z_d, \hat s)\right)$ vanishes, while for  $m_{l^+l^-}\sim m_c$ it is at a level of few percent! Using \cite{Burdman:2001tf,Fajfer:2001sa} $C_9^{\text{IL}}$ \eqref{C9IL} instead of $\hat{C}_9^{\text{eff}}$ \eqref{hatc9}, which includes the QCD corrections, one overestimates the $d\text{Br}(c\to l^+l^-)/d\hat s$.

Explicitly, the branching ratio is \cite{Fajfer:2001sa}
\begin{equation}
\begin{split}
\frac{\text{Br}(c\to ul^+l^-)}{d\hat{s}}=&\frac{G_F^2 \alpha_{\text{QED}}^2 m_c^5}{768 \pi^5\Gamma(D^0)}|V_{cb}^* V_{ub}|^2 (1-\hat s)^2\bigg[4\left(1+\frac{2}{\hat s}\right) |\hat{C}_7^{\text{eff}}|^2\\
&+\frac{1+2\hat s}{16}\left(|\hat{C}_9^{\text{eff}}|^2+|\hat{C}_{10}^{\text{eff}}|^2\right)+3 \Re\left(\hat{C}_7^{\text{eff}*}\hat{C}_9^{\text{eff}}\right)\bigg],
\end{split}
\end{equation}
where we write $\hat s=(m_{l^+l^-}/m_c)^2$ as  before. For the value of $\hat{C}_7^{\text{eff}}$ we use the two-loop result of Ref. \cite{Greub:1996wn}, $\hat{C}_7^{\text{eff}}=\lambda_s (0.007+0.020 i)(1\pm 0.2)$, with $\lambda_s$ defined after Eq. \eqref{C9IL}. The dominant contribution to $\hat{C}_7^{\text{eff}}$ comes from the insertion of $Q_2^q$ operator, while the contributions from the insertion of $Q_1^q$ operators vanish because of color structure. The coefficient $\hat{C}_{10}^{\text{eff}}\simeq 0$ in the Standard Model.

Using $m_c=1.4$ GeV   one arrives at
\begin{equation}
\begin{split}
\text{Br}(c\to u e^+e^-)&=2.4 \times 10^{-10},\\
\text{Br}(c\to u \mu^+\mu^-)&=0.5 \times 10^{-10},
\end{split}
\end{equation}
where the dominant contribution comes from the $\hat{C}_7^{\text{eff}}$ part of the amplitude. This is in contrast to Refs. \cite{Burdman:2001tf,Fajfer:2001sa}, where $\hat{C}_9^{\text{eff}}$ was estimated using  $C_9^{\text{IL}}$. This lead to  the branching ratios of one (for $e^+e^-$) to two  (for $\mu^+ \mu^-$) orders of magnitude higher, with $\hat{C}_9^{\text{eff}}$ contribution dominating the branching ratio.

The suppression of QCD corrected $\hat{C}_9^{\text{eff}}$ \eqref{hatc9} compared to  $C_9^{\text{IL}}$ \eqref{C9IL} comes from two sources. The cancellation of $s$ and $d$ quark contributions in \eqref{hatc9} is very strong even at moderate values of $\hat{s}$, with  $\left( h(z_s,\hat s)-h(z_d, \hat s)\right)\le 10\% $ for  $\hat{s}\ge 0.3$. There is also a sizable cancelation between $C_1(m_c)$ and $C_2(m_c)$ in \eqref{hatc9}. This cancelations could in principle be modified by the two-loop QCD corrections to $Q_{1,2}$ matrix elements\footnote{The existing two-loop calculations of $Q_{1,2}$ matrix elements in $b\to s l^+l^-$ \cite{Asatrian:2001de,Asatryan:2001zw} have been done for small $\hat s$, where no substantial increase in $c\to u l^+l^-$ is expected.}. If the cancelations were completely lifted, one can estimate the possible effect by   $\hat{C}_9^{\text{eff}}\sim \alpha_s(m_c) C_9^{\text{IL}}$. This  leads to roughly the same prediction for $\text{Br}(c\to u e^+e^-)$, while it can increase $\text{Br}(c\to u \mu^+\mu^-)$, as   $\hat{C}_9^{\text{eff}}$ affects mostly the higher $\hat s$ part of the decay width distribution.

Note, that the  calculation of $c\to u l^+ l^-$ is in many respects  different than the calculation of $b\to s l^+ l^-$. The  operators $Q_{1,2}^{u,c(b\to s)}$  in $b\to s l^+ l^-$ are equivalent to  $Q_{1,2}^{d,s}$ operators in the $c\to u l^+l^-$ transition, but with different CKM factors multiplying the operators in the effective Lagrangian. In $b\to s l^+ l^-$ then  only the $Q_{1,2}^{c(b\to s)}$ operators contribute, as the contributions coming from the $Q_{1,2}^{u(b\to s)}$ operators are  $V_{ub}$ suppressed. Hence, there is no approximate cancellation of the type $\left( h(z_s,\hat s)-h(z_d, \hat s)\right)$ found above.  Note also, that in $b\to s l^+ l^-$ the penguin operators $Q_{3,\dots,10}$ are not CKM suppressed relative to $Q_{1,2}$ and have to be taken into account, contrary to the $c\to u l^+ l^-$ case, where penguin operators are $V_{ub}$ suppressed.

The $V_{ub}$ suppression of penguin operators $Q_{3,\dots,10}$ is of course present in the calculation of exclusive charm decays, where the insertions of $Q_{1,2}$ operators again dominate the rate. This will be discussed in more detail for the case of $D^0 \to l^+l^- \gamma$ decay in the following section. Before we proceed with the calculation, let us mention the commonly used terminology of long distance (LD) and short distance (SD) contributions. These  are usually separated in the  discussion of weak radiative decays
$q' \to q \gamma \gamma$ or $q' \to q \gamma $ decays.
The SD contribution in these transitions is a  result of the
penguin-like transition induced by the operators $Q_{7,9,10}$, while the long distance contribution
arises  from insertions of $Q_{1,2}$ operators, when the off- or on-shell photon is emitted
from the
quark legs. We will follow this classification in the following.

\begin{figure}
\begin{center}
\epsfig{file=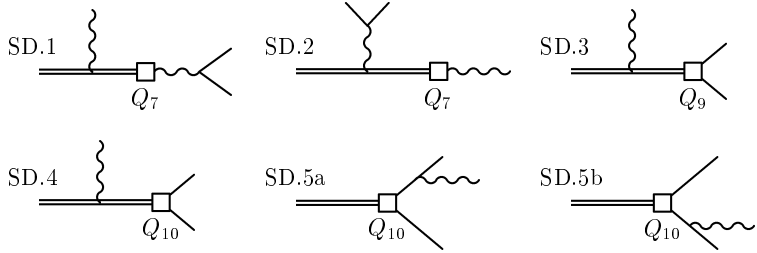, height=3.5cm}
\caption{\footnotesize{The short distance diagrams. The effective weak Lagrangian vertex is denoted by an empty square. The relevant operator is denoted as well. }}\label{SDdiagr}
\end{center}
\end{figure}

At this point, we mention our result for the SD contribution coming
from the operators $Q_{7,9,10}$ (see Fig. \ref{SDdiagr}). This contribution turns out indeed to be very small in the SM, due to the CKM suppression \eqref{weak-eff-lagr}. Evaluating
the expectation values of operators $Q_{7,9,10}$ by using heavy quark
symmetry as described in Eq. \eqref{eq-105} of the next sub-section (see the explicit
expressions in Appendix \ref{app-C}) and using the values of Wilson coefficients 
listed in Table  \ref{tab-Wilson} and in Eq. \eqref{C7numbers}, one arrives at the corresponding branching ratios for
$D^0\to l^+l^-\gamma$ of $10^{-17}-10^{-18}$. This is negligible compared to the LD contributions calculated in the subsections \ref{nonresonant}, \ref{resonant}. Our result for the
SD contribution to these decays is several orders of magnitude smaller
than the result of \cite{Geng:2000if}, which was obtained with an unrealistic value of $C_9$ that did not include QCD corrections.

 \subsection{Nonresonant LD contributions}
\label{nonresonant}

Turning now  to the LD contributions, we start with the nonresonant contributions.  As we will see later on, it is in the nonresonant contributions that the extensions of Standard Model can show up.

The most general invariant amplitude for $D^0 \to l^+l^- \gamma $ decay following  from the effective Lagrangian \eqref{weak-eff-lagr} is\footnote{We use $\epsilon^{0123}=1$.}
\begin{equation}
\begin{split}
M &=M_0^{\mu\nu}\epsilon_\mu^*(k)\frac{1}{p^2} \bar{u}(p_1)\gamma_\nu v(p_2)+M_5^{\mu\nu}\epsilon_\mu^*(k)\frac{1}{p^2} \bar{u}(p_1)\gamma_\nu\gamma_5 v(p_2)+\\
&\qquad+ M_{\text{BS}}(p^2)\Big[ \bar{u}(p_1)\Big(\frac{\sls{\epsilon}^* \sls{p}_D}{p_1 \cdot k}-\frac{\sls{p}_D \sls{\epsilon}^*}{p_2 \cdot k}\Big) \gamma_5 v(p_2)\big],
\end{split} \label{inv_ampl}
\end{equation}
where
\begin{equation}
M_{0,5}^{\mu\nu}=C_{0,5}(p^2) \big(\eta^{\mu\nu}-\frac{p^\mu k^\nu}{p\negcdot k}\big)+ D_{0,5}(p^2) \epsilon^{\mu \nu \alpha \beta} k_\alpha p_\beta, \label{inv_amp}
\end{equation}
with $p_{1,2}$ the four-momenta of lepton and antilepton respectively, $p=p_1+p_2$ the momentum of lepton pair, $k$ the photon momentum and $\epsilon_\mu$ its polarization vector. The form factors $C_{0,5}(p^2)$, $D_{0,5}(p^2)$, $M_{\text{BS}}(p^2)$ are functions of $p^2$ only and in particular do not depend on $k\cdot p_1$ or $k\cdot p_2$. $C_0,D_5$ are parity violating terms, while
$C_5,D_0$ and the bremsstrahlung part of the amplitude, $M_{\text{BS}}$,  are parity conserving.

The partial decay width is then
\begin{equation}
\begin{split}
\frac{d\Gamma}{dp^2}=&\frac{1}{16 \pi^3 m_D^3} \Bigg\{ \frac{k \negcdot p}{3 p^2}\sqrt{1-4\hat{\mu}_p^2}\bigg[\big( |C_0|^2+|D_0|^2(k\negcdot p)^2\big) (1+2\hat{\mu}_p^2)+\\
&+\big( |C_{5}|^2+|D_{5}|^2(k\negcdot p)^2\big) (1-4\hat{\mu}_p^2)\bigg]\\
&+\frac{|M_{\text{BS}}|^2}{k\negcdot p} \Big[\left((p^2)^2+m_D^2(m_D^2-4 m^2)\right) \; \ln\Big(\frac{1+\sqrt{\quad}}{1-\sqrt{\quad}}\Big)-2 p^2 m_D^2\sqrt{\quad}\Big]\\
&+4 \Im(D_0 M_{\text{BS}}^*) \frac{m}{p^2} (k\negcdot p)^2 \; \ln\Big(\frac{1+\sqrt{\quad}}{1-\sqrt{\quad}}\Big)\Bigg\},
\end{split}
\end{equation}
where  $\hat{\mu}_p^2=m^2/p^2$, with $m$ the lepton mass, $\sqrt{\quad}=\sqrt{1-4 \hat{\mu}_p^2}$, while $k\negcdot p=(m_D^2-p^2)/2$. We checked that this expression agrees
with the similar expression  for the partial decay width  $K_L\to l^+l^-  \gamma$  as given in \cite{D'Ambrosio:1994ae,D'Ambrosio:1996sw}, as well as with the $B\to l^+l^-\gamma$ decay width as given in \cite{Aliev:2001bw}. 


\begin{figure}
\begin{center}
\epsfig{file=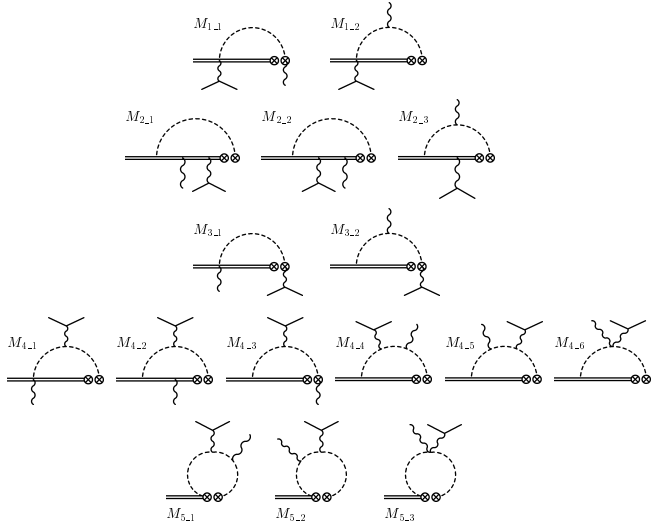, height=7cm}
\caption{\footnotesize{Nonvanishing one loop diagrams. The dashed lines
represent charged Goldstone bosons flowing in the loop ($K^+,\pi^+$), while the double lines represent  heavy mesons, $D$ and $D^*$. The two crossed circles denote weak vertex calculated in the factorization approximation. The sum of diagrams in each row is gauge invariant and finite.}}\label{fig-1}
\end{center}
\end{figure}

The  nonresonant LD contributions will arise in our approach from chiral loop  contributions  shown on Figure \ref{fig-1}. The weak vertices receive contributions from $Q_{1,2}$ operators in the effective Lagrangian \eqref{weak-eff-lagr}. The sizes of these contributions are estimated using factorization approximation.  The effective \cite{Bauer:1986bm} four quark
nonleptonic $\Delta C = 1$
  weak Lagrangian
 is then
\begin{equation}
{\cal L}=-\frac{G_F}{\sqrt{2}} \sum_{q=d,s}V_{uq}
 V_{cq}^* \big[ a_1  \big(\bar{q} \Gamma^\mu c) (\bar{u}
\Gamma_\mu q)+
 a_2 (\bar{u}\Gamma^\mu c) (\bar{q} \Gamma_\mu q)\big],
\label{eq-107}
\end{equation}
where $\Gamma^\mu=\gamma^\mu(1-\gamma_5)$,   $a_i$ are effective
Wilson
coefficients, $V_{q_i q_j}$ the  CKM matrix
elements, while products of  currents in \eqref{eq-107} are understood to be evaluated in the factorization approximation. We use the phenomenologically motivated values \footnote{The new factorization values of effective Wilson coefficients correspond to $N_c\to \infty$ limit and are in terms of Wilson coefficients $a_1=C_2$, $a_2=C_1$.}  $a_1=1.26$, $a_2=-0.49$ of ``new factorization'' \cite{Buras:1994ij}.  It is worth pointing out that long distance interactions will
contribute only if the $SU(3)$
flavor symmetry  is broken, i.e. if $m_s\neq m_d$. Namely, due to
$V_{ud}V_{cd}^* \simeq -V_{us}V_{cs}^*$, if  $m_d = m_s$
the contributions arising from the weak Lagrangian (\ref{eq-107})
cancel. Note also that in diagrams of Fig. \ref{fig-1} only the term proportional to $a_1$ contributes. The $a_2$ part of effective Lagrangian \eqref{eq-107} gives rise to the resonant LD contributions and will be discussed in the next section.

We calculate the nonresonant LD contributions in the framework of Heavy Quark Chiral Perturbation Theory  HQ$\chi$PT \cite{Fajfer:2001ad,Casalbuoni:1996pg}. 
This model will serve us  when hadronizing the currents \cite{Wise:hn}
of the quark  effective weak
Lagrangian. In the framework of HQ$\chi$PT  a number of coupling constants appear that are fixed from experiment as discussed in Ref. \cite{Fajfer:2001ad} and are listed in Table \ref{table-const}. In the following we first give a short introduction to HQ$\chi$PT and then turn to the discussion of results.

In  the leading order of HQ$\chi$PT the light
pseudoscalar mesons are described by 
the  usual ${\cal O}(p^2)$ chiral Lagrangian 
\begin{equation}
{\cal L}_{\text{str}}^{(2)}=\frac{f^2}{8} \tr (\partial^\mu \Sigma
\partial_\mu \Sigma^\dagger)+\frac{f^2 B_0}{4}\tr({\cal M}_q \Sigma
+{\cal M}_q\Sigma^\dagger) \; ,\label{eq-11}
\end{equation}
where $\Sigma = \exp{(2 i \Pi/f)}$ with
$\Pi =  \sum_j \frac{1}{\sqrt{2}}\lambda^j \pi^j$ containing the
Goldstone
bosons $\pi,
K,
\eta$, while the trace $\tr$ runs over flavor indices and ${\cal
M}_q=\diag
(m_u,m_d,m_s)$ is the current quark mass matrix. From this Lagrangian
 we can deduce the
 light weak current of the order ${\cal O}(p)$
\begin{equation}
j_\mu^a \, = \, -i\frac{f^2}{4}\tr(\Sigma \partial_\mu
\Sigma^\dagger\lambda^a) \; ,
\label{jX}
\end{equation}
corresponding to the quark current
$j_\mu^a=\bar{q}_{L}\gamma_\mu\lambda^aq_{L}$ (with $\lambda^a$ an SU(3)
flavor matrix).

For the heavy mesons interacting with light pseudoscalars  we have
the following lowest order ${\cal O}(p)$  chiral Lagrangian
\begin{equation}
{\cal L}_{\text{str}}^{(1)}=-\Tr(\bar{H}_{a}iv\negcdot D_{ab}H_{b})+g
\Tr(\bar{H}_{a}H_{b} \gamma_\mu {\cal A}_{ba}^\mu \, \gamma_5) \; ,
\label{eq-8}
\end{equation}
where $ D_{ab}^\mu H_b=\partial^\mu H_a - H_b{\cal V}_{ba}^\mu$,
 while the trace $\Tr$ runs over Dirac indices.
Note that in (\ref{eq-8}) and the rest of this section
$a$ and $b$ are {\em flavor} indices. The vector and axial vector fields
${\cal V}_{\mu}$ and
${\cal A}_\mu$ in (\ref{eq-8}) are given by:
\begin{equation}
{\cal V}_{\mu} = \frac{1}{2}(\xi\partial_\mu\xi^\dagger
+\xi^\dagger\partial_\mu\xi), \qquad   \qquad
{\cal A}_\mu = \frac{i}{2}
(\xi^\dagger \partial_\mu\xi -\xi\partial_\mu\xi^\dagger) \; ,
\label{defVA}
\end{equation}
where $\xi = \exp{(i \Pi/f)}$. The heavy meson field
 $H_{a}$ contains
 a spin zero $P_a$ and spin one $P_{a\mu}$ boson fields
\begin{equation}
H_{a} = P_+ (P_{\mu a} \gamma^\mu -
 P_{ a} \gamma_5),\qquad
\overline{H}_{a}=\gamma^0 (H_{a})^\dagger \gamma^0,
\label{barH}
\end{equation}
with $ P_{\pm}=(1 \pm \sls{v})/2 $ the  projection
operators.

From symmetry grounds,
 the heavy-light weak current is  bosonized
in the following way \cite{Wise:hn}
\begin{equation}
\overline{q}_a \gamma^\mu P_L Q =  \frac{i\alpha}{2} \Tr [\gamma^\mu \, P_L \, H_{b}\,
\xi_{ba}^\dagger]
\; ,
\label{JH}
\end{equation}
where $P_{R, L}=(1\pm\gamma_5)/2$,  $Q$ is the heavy quark field in the
 full theory, in our case a $c$-quark field, and $q$ is the light quark
 field. Note that the current (\ref{JH}) is  ${\cal O}(p^0)$ in the chiral
counting. The constant  $\alpha$ is related to the
 physical decay constant $f_D$ through the well known matrix
 element
\begin{equation}
\langle  0| \overline{u}\gamma^\mu \gamma_5 c|D^0 \rangle
=ip_D^\mu f_D \;,
\label{fD}
\end{equation}
from which $\alpha=f_D\sqrt{m_D}$. From \cite{Hagiwara:pw} one deduces $f_{D_s}=268\pm25$ GeV and $\alpha=0.38\pm0.04 \; \text{GeV}^{3/2}$. 
 In the same way as the heavy-light current (\ref{JH}),  operators of more general structure  $ (\bar{u}\Gamma c)$, with $\Gamma$ an arbitrary product of Dirac matrices,  can be translated into
an operator containing meson fields only \cite{Falk:1993fr}
\begin{equation}
\big(\bar{u} \Gamma c\big)\to \frac{i
\alpha}{2}\Tr[P_R \Gamma H_b\xi_{ba}^\dagger] + \frac{i
\alpha}{2}\Tr[P_L \Gamma H_b\xi_{ba}].\label{eq-105}
\end{equation}

\begin{table}[h]
\begin{center}
\begin{tabular}{|c|c||c|c|}\hline
$f$&$132$ MeV& $g$&$0.59\pm0.08$\\ \hline
$\alpha$&$0.38\pm0.04 \;\text{GeV}^{3/2}$& $\beta$&$2.3\pm0.2\; \text{GeV}^{-1}$\\ \hline

$a_1$&$1.26$ &$a_2$&$-0.49$ \\ \hline
\end{tabular}
\caption{\footnotesize{Coupling constants appearing in HQ$\chi$PT, that is used in the estimates of nonresonant contributions. For further details see text and \cite{Fajfer:2001ad}. In the last row values of effective Wilson coefficients are given \cite{Buras:1994ij}. Loop integrals are calculated in $\overline{\text{MS}}$ scheme with scale $\mu=1\;\text{GeV}$, while in \eqref{eq-100} $m_c=1.4$ GeV.} }
\label{table-const}
\end{center}
\end{table}

The photon couplings  are obtained by gauging the Lagrangians
\eqref{eq-11}, \eqref{eq-8} and the light current \eqref{jX}
with the $U(1)$ photon field $B_\mu$.
The covariant derivatives are
then ${\cal D}_{ab}^\mu H_b=
\partial^\mu H_a +i e B^\mu (Q'H-H Q)_a-H_b {\cal V}_{ba}^\mu$
and
${\cal D}_\mu\xi =\partial_\mu \xi +
i e B_\mu [Q,\xi]$ with $Q=\diag (\frac{2}{3},-\frac{1}{3},
-\frac{1}{3})$
and $Q'=\frac{2}{3}$ (for the  case of $c$ quark). The vector and
axial vector
fields  \eqref{defVA} and the light weak current \eqref{jX} contain after gauging the covariant
derivative ${\cal D}_\mu$ instead
of $\partial_\mu$. However, the gauging procedure alone does not
introduce
a transition  $DD^*\gamma$ without emission or absorption
of
additional Goldstone boson. To
describe
this
electromagnetic interaction
we follow \cite{Stewart:1998ke} introducing an additional gauge invariant
contact
term with
coupling $\beta$
of dimension -1
\begin{equation}
{\cal  L}_\beta=-\frac{\beta e}{4} \Tr \bar{H}_a H_b \sigma^{\mu
\nu}F_{\mu
\nu} Q_{ba}^\xi -\frac{e}{4 m_Q} Q' \Tr \bar{H}_a
\sigma^{\mu \nu} H_a
F_{\mu\nu},\label{eq-100}
\end{equation}
where $Q^\xi=\frac{1}{2}(\xi^\dagger Q \xi+\xi Q \xi^\dagger)$ and
$F_{\mu\nu}=\partial_\mu B_\nu- \partial_\nu B_\mu$.
The first term concerns the contribution
of the light quarks in
the heavy meson and the  second term
describes  emission of
a photon
from the heavy quark. Its coefficient  is fixed by heavy quark
symmetry. From this "anomalous" interaction, both $H^* H \gamma$ and
$H^*H^*\gamma$ interaction terms arise. Even though the
Lagrangian \eqref{eq-100} is formally $1/m_Q \sim m_q$ suppressed,
we
do not neglect it completely. We do not take it into account in chiral loop contributions of Fig. \ref{fig-1}, as  it has been found to give a rather small  contribution in a very similar case of $D^0\to \gamma\gamma$ analysis \cite{Fajfer:2001ad}. The $D^0\to D^{0*}\gamma$ transition will be, however, needed to estimate the  short distance contributions shown on Fig  \ref{SDdiagr}. These will give numerically irrelevant contributions for SM predictions but will be important later on, when we extend the analysis to MSSM case. Note also, that the Lagrangian \eqref{eq-100}
in principle
receives a number of other contributions at the order of $1/m_Q$.
However, these
can be absorbed in the definition of $\beta$ for the processes
considered \cite{Stewart:1998ke}.

Using HQ$\chi$PT as described above, one arrives at the set of nonzero ${\cal O }(p^3)$ diagrams listed in Fig. \ref{fig-1}.  Each row of diagrams on Fig \ref{fig-1} is a gauge invariant set. The sum of diagrams in each row is also finite. Separate diagrams are in general divergent and are regulated using dimensional regularization. Further details on this subject can be found in the  appendix  \ref{app-A}. The explicit expressions of the corresponding amplitudes can be found in appendix \ref{app-B}. Note that the chiral loop contributions of Fig. \ref{fig-1} contribute only to the $M_0^{\mu\nu}$ part of the invariant amplitude \eqref{inv_ampl} . Namely, the $l^+l^-$ pair couples to the charged mesons  in the loop only via electromagnetic current. This also leads to the $1/p^2$ photon pole in the amplitude ($p$ being the momentum of the lepton pair). The LD nonresonant contributions coming from Fig. \ref{fig-1} thus exhibit a pole behaviour at small lepton momenta. This pole is either cut off by the phase space because of nonzero lepton masses  ($p^2=4 m^2$), or by  experimental limitations   due to Dalitz conversion \cite{Burdman:2001tf}.

\begin{figure}
\begin{center}
\epsfig{file=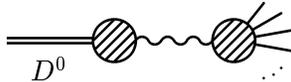}
\caption{\footnotesize{One particle reducible diagrams with photon connecting initial (pseudo)scalar and final state particles are zero.}}\label{TwoBlobs}
\end{center}
\end{figure}

Note  that there is no  photon bremsstrahlung off the final lepton pair in the chiral loop contributions. Namely, diagrams of the type shown on Fig. \ref{TwoBlobs},  with initial meson being a (pseudo)scalar, and with a photon connecting the two blobs, vanish  due to gauge invariance. 

The diagrams of Fig. \ref{fig-1} are  evaluated in the minimal subtraction ($\overline{\text{MS}}$) renormalization scheme. However, the sum of diagrams is finite and scheme independent. We use the values of coupling constants  listed in Table \ref{table-const}. Integrating over the whole available phase space  one arrives at the estimates
\begin{equation}
\text{Br}(D^0\to e^+e^-\gamma)_{\text{nonres}}=1.29 \times 10^{-10}, \qquad \text{Br}(D^0\to \mu^+\mu^-\gamma)_{\text{nonres}}=0.21 \times 10^{-10}.\label{nonres-res}
\end{equation}
Due to a photon pole,  the larger part of the electron channel branching ratio comes from the region of the phase space with $p^2\sim 0$. The phase space is cut off by muon masses at much higher $p^2$, giving a smaller contribution of nonresonant LD effects to this decay channel.

\subsection{Resonant LD contributions}\label{resonant}
\begin{figure}
\begin{center}
\epsfig{file=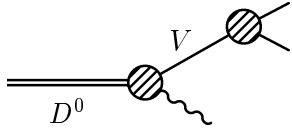}
\caption{\footnotesize{The mechanism of $D^0\to l^+l^-\gamma$ decay through intermediate vector resonance state $V$.}}\label{Reson}
\end{center}
\end{figure}

The mechanism of the decay $D^0\to  l^+l^-\gamma$ through resonant intermediate state is depicted on Fig \ref{Reson}. The $D^0$ meson first decays into a vector meson and  a photon, $D^0\to V \gamma$. The vector meson than decays into a lepton pair, completing the cascade $D^0\to V\gamma\to l^+l^- \gamma$. The decay width coming from this mechanism can be written as \cite{Lichard:1998ht}
\begin{equation}
\frac{d \Gamma_{D^0\to V \gamma\to l^+ l^- \gamma}}{d p^2}=\Gamma_{D^0\to V\gamma} \frac{1}{\pi}\frac{\sqrt{p^2}\; }{(M_V^2-p^2)^2+M_V^2 \Gamma^2}\Gamma_{V\to l \bar{l}}, \label{factor}
\end{equation}
where $p$ is the momentum of the lepton pair, while $M_V$ and $\Gamma$ are the mass and the decay width of the vector meson resonance.  Several assumptions go into the  derivation of the simple, but physically well motivated  formula \eqref{factor}. First of all the interference with other channels is neglected. Under this approximation the formula  is generally valid for the case of scalar resonances.  Following the reasoning of Ref. \cite{Lichard:1998ht} it is easy to show, that Eq. \eqref{factor} is valid also for the case of electromagnetic decay of vector resonance into a lepton pair.

Since vector resonances $\rho, \omega, \phi$  are relatively narrow Eq. \eqref{factor} can be further simplified using the narrow width approximation $\Gamma\ll M_V$
\begin{equation}
\text{Br}(D^0\to V\gamma\to l^+ l^- \gamma)= \text{Br}(D^0\to V \gamma) \text{Br}(V\to l^+l^-). \label{narrow_width}
\end{equation}
The narrow width approximation is valid at $5\%$ level for $\rho$, and below $1\%$ for $\omega, \phi$ mesons. To obtain numerical estimates,  the experimental data on the branching ratios $\text{Br}(V\to l^+l^-)$ \cite{Hagiwara:pw} can be used. On the other hand none of the decays  $D^0\to V\gamma$ have been measured yet.  We thus use the theoretical predictions  of branching ratios $\text{Br}(D^0\to V \gamma)$. As the central values  we use the recent predictions of Ref. \cite{Prelovsek:2000rj}, where a reanalysis of Ref. \cite{Fajfer:1998dv} has been performed using the quark model to determine relative
phase uncertainties. As a comparison we also list in Table \ref{table-input2} the  predictions of Ref. \cite{Burdman:1995te}. Note that for the upper limit predictions in \cite{Burdman:1995te} VMD model was used, with the main numerical input the experimental value of  $\text{Br}(D^0\to \rho^0 \phi)$. However,  the central value of this branching fraction as cited in \cite{Hagiwara:pw}  has decreased by a factor of three between 1994-2002. Thus the upper limits on predictions of  \cite{Burdman:1995te} should be divided by three, bringing the values in fair agreement with \cite{Prelovsek:2000rj}.

\begin{table}[h]
\begin{center}
\begin{tabular}{|c|r||c|r|}\hline
Decay& Exp.  \cite{Hagiwara:pw} & Decay & Exp. \cite{Hagiwara:pw}\\ \hline\hline
$\text{Br}(\rho^0\to e^+e^-)$ & $(4.54\pm0.10)\times 10^{-5}$ & $\text{Br}(\rho^0\to \mu^+\mu^-)$ & $(4.60\pm0.28)\times 10^{-5}$\\ \hline
$\text{Br}(\omega \to e^+e^-)$ & $(6.95\pm0.15)\times 10^{-5}$ & $\text{Br}(\omega \to \mu^+\mu^-)$ & $(9.0\pm 3.1)\times 10^{-5}$\\ \hline
$\text{Br}(\phi\to e^+e^-)$ & $(2.96\pm 0.04)\times 10^{-4}$ & $\text{Br}(\phi\to \mu^+\mu^-)$ & $(2.87\genfrac{}{}{0pt}{}{+0.18}{-0.22})\times 10^{-4}$\\\hline
\end{tabular}
\caption{\footnotesize{Branching ratios of vector mesons decaying to a lepton pair as compiled in Ref. \cite{Hagiwara:pw}. }}
\label{table-input1}
\end{center}
\end{table}

\begin{table}[h]
\begin{center}
\begin{tabular}{|c|c|c|c|}\hline
 Decay & Theor.  \cite{Prelovsek:2000rj}& Theor. \cite{Burdman:1995te} & Exp. \cite{Hagiwara:pw}\\ \hline\hline
 $\text{Br}(D^0\to \rho^0 \gamma)$ & $1.2\times 10^{-6} $ & $(1 -5)\times 10^{-6} $ & $ <2.4 \times 10^{-4}$\\ \hline
 $\text{Br}(D^0\to \omega \gamma)$ & $1.2\times 10^{-6} $ & $ \simeq2 \times 10^{-6} $& $ <2.4 \times 10^{-4}$\\ \hline
 $\text{Br}(D^0\to \phi \gamma)$ & $3.3 \times 10^{-6} $& $(1-34) \times 10^{-6} $ & $ <1.9 \times 10^{-4}$\\ \hline
\end{tabular}
\caption{\footnotesize{Theoretical predictions for decays $D^0\to V\gamma$  \cite{Prelovsek:2000rj,Burdman:1995te}. Predictions of Ref. \cite{Prelovsek:2000rj} are used as central values (see also comments in text). 
In the last column the experimental upper limits are listed.}}
\label{table-input2}
\end{center}
\end{table}

Using the values compiled in Tables \ref{table-input1}, \ref{table-input2} together with Eq. \eqref{narrow_width} one immediately arrives at 
\begin{align}
\text{Br}(&D^0\to\rho\gamma\to l^+l^- \gamma) \sim 5 \times 10^{-11} ,\label{reson1}\\
\text{Br}(&D^0\to\omega\gamma\to l^+l^- \gamma) \sim 8 \times 10^{-11},\\
\text{Br}(&D^0\to\phi\gamma\to l^+l^- \gamma) \sim 10^{-9} ,\label{reson3}
\end{align}
 with $l^+l^-=e^+e^-, \mu^+\mu^-$. Above  we have used the fact that differences between $e^+e^-$ and $\mu^+\mu^-$ decay modes in the standard model come from the phase space differences only. These are relatively small compared to other theoretical and experimental uncertainties entering predictions \eqref{reson1}-\eqref{reson3}, and are as such neglected.

\begin{figure}
\begin{center}
\epsfig{file=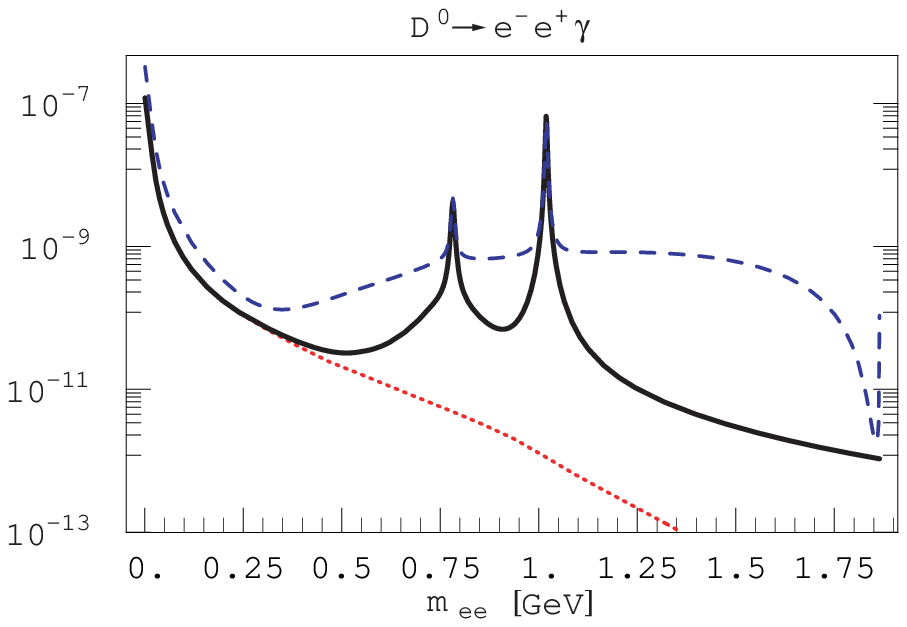, height=5.4cm}
\epsfig{file=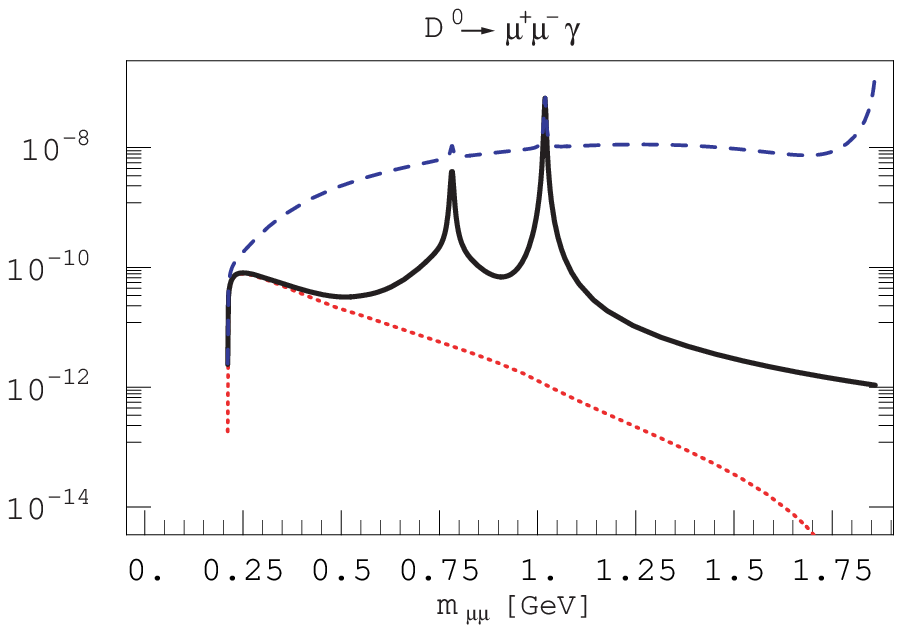, height=5.4cm}
\caption{\footnotesize{The normalized decay width distribution $(d\Gamma/d p^2)/\Gamma$ as function of effective lepton pair mass $m_{l^+ l^-}$ (where $m_{l^+ l^-}^2=p^2$) for $e^+e^-$ (left plot)  and $\mu^+\mu^-$ (right plot)  final lepton pair. The  dotted  line denotes SM nonresonant contribution, solid black line denotes full SM prediction, while dashed line denotes largest possible MSSM contribution with $R$ parity violation.}}\label{fig-res}
\end{center}
\end{figure}

As seen from the estimates \eqref{reson1}-\eqref{reson3} the largest contribution to $D^0\to l^+l^-\gamma$ comes from the intermediate $\phi$ resonance, being approximately one order of magnitude larger than the other two contributions.  Note also, that in the region of $p^2$, where vector resonances are important, the nonresonant contribution calculated in the previous section is several orders of magnitude smaller. We can thus safely neglect possible interference between nonresonant and resonant contributions and simply add resonant contributions \eqref{reson1}-\eqref{reson3} to the nonresonant ones \eqref{nonres-res}. The decay width distribution is plotted on Fig. \ref{fig-res}, while the predicted branching ratios are
\begin{equation}
\text{Br}(D^0\to e^+e^-\gamma)_{\text{SM}}=1.2 \times 10^{-9}, \qquad \text{Br}(D^0\to \mu^+\mu^-\gamma)_{\text{SM}}=1.1 \times 10^{-9}.  \label{SM-pred}
\end{equation}
Note that if the values of Ref. \cite{Burdman:2001tf} had been used, the predicted branching ratios could be utmost a factor of three higher.

Incidentally Fig. \ref{fig-res} also explains, why the $D^0\to V\gamma\to \gamma^*\gamma$ cascade could be  neglected in the $D^0\to \gamma\gamma$ decay rate calculation of Ref. \cite{Fajfer:2001ad}. Namely, for $\gamma^*$ almost on-shell the decay width is dominated by the nonresonant contributions. In the calculation of $D\to \gamma\gamma$ \cite{Fajfer:2001ad} these were described using HQ$\chi$PT along the lines presented in section \ref{nonresonant}.

\section{Beyond the Standard Model}
In this section we will consider possible effects of physics beyond the Standard Model, that could enhance the predicted branching ratios \eqref{SM-pred}. The effects of new physics show up in the models we considered in the values of Wilson coefficients
\begin{equation}
C_i^{\text{new}}=C_i+\delta C_i,
\end{equation}
where $C_i$ are the SM values of Wilson coefficients listed in Table \ref{tab-Wilson} and in Eq. \eqref{C7numbers}, while $\delta C_i$ denote the changes due to new physics effects. Note that the general feature of all the SM extensions is to overcome the $V_{cb}^* V_{ub}$ suppression of penguin operators $Q_{7,9,10}$ \eqref{weak-eff-lagr}. Another general feature is that the new physics effects will extend the basis of penguin operator \eqref{list-oper} by operators $Q_{7,9,10}'$ with quark chiralities switched (i.e. they are obtained by exchanging $P_R\leftrightarrow P_L$ in \eqref{Q7},\eqref{Q9}).

\subsection{Minimal Supersymmetric Standard Model}
We start with the simplest supersymmetric extension of SM, the Minimal Supersymmetric Standard Model (MSSM). It is constructed by putting the SM fermions in chiral multiplets and the SM gauge bosons in the vector multiplets, thus in effect doubling the spectrum of Standard Model fields. If no particular SUSY breaking mechanism is assumed the MSSM Lagrangian contains well over 100 unknown parameters. It is thus very useful to adopt the so-called mass insertion approximation. In this approximation the basis of fermion and sfermion states is chosen such that all the couplings of these particles to neutral gauginos are flavor diagonal, but then the squark mass matrices are not diagonal. The squark propagators are then expanded in terms of nondiagonal elements, where  mass insertions  induce changes of squark flavor \cite{Hall:1985dx}. The mass insertions are parametrized as
\begin{equation}
(\delta^u_{ij})_{AB}=\frac{(M_{ij}^u)^2_{AB}}{M_{\tilde{q}}^2},
\end{equation}
where $i \ne j$ are flavor indices, $A,B$ denote chirality, $(M_{ij}^u)^2$ are the off-diagonal elements of up-type squark mass matrices and $M_{\tilde{q}}$ is the average squark mass.

The largest contribution to $c\to ul^+l^-$ transition is expected from gluino-squark exchanges \cite{Burdman:2001tf,Fajfer:2001sa,Lunghi:1999uk}. Allowing for only one insertion, the contributions from gluino-squark exchange diagrams are
\begin{subequations}\label{gluino}
\begin{align}
V_{cb}^* V_{ub} \delta C_7 &= \frac{8}{9} \frac{\sqrt{2}}{G_F M_{\tilde q}^2} \pi \alpha_s \left[(\delta_{12}^u)_{LL} \frac{P_{132}(z)}{4}+ (\delta^u_{12})_{LR} P_{122}(z) \frac{M_{\tilde{g}}}{m_c} \right],\\
V_{cb}^* V_{ub} \delta C_9&=\frac{32}{27} \frac{\sqrt{2}}{G_F M_{\tilde{q}}^2}\pi \alpha_s (\delta_{12}^u)_{LL} P_{042}(z),\\
 V_{cb}^* V_{ub} \delta C_{10}&\simeq 0,
\end{align}
\end{subequations}
where $z=M_{\tilde g}^2/M_{\tilde q}^2$, while the functions $P_{ijk}(z)$ are
\begin{equation}
P_{ijk}(z)=\int_0^1 dx \frac{x^i (1-x)^j}{(1-x+z x)^k}.
\end{equation}
The Wilson coefficient $C_{10}$ receives first nonzero contributions from double mass insertions, therefore we neglect it in the following. The Wilson coefficients $C_{7, 9, 10}'$ corresponding to the operators with ``wrong chirality''  receive contributions from gluino-squark exchanges that are of the same form as expressions \eqref{gluino}, but with the interchange $L\leftrightarrow R$.

In numerical evaluation of possible MSSM effects we use gluino mass $M_{\tilde g}=250$ GeV and the average squark mass $M_{\tilde q}=250$ GeV, that are given by the lower experimental bounds \cite{Hagiwara:pw}. For the bounds on the mass insertions we use the analysis of  \cite{Fajfer:2001sa,Prelovsek:2000xy}. The strongest bounds on mass insertion parameters $(\delta^u_{12})_{LR}$ are obtained by requiring that the minima of the scalar potential do not break charge or color, and that they are bounded from below \cite{Prelovsek:2000xy,Casas:1996de}, giving
\begin{equation}
|(\delta^u_{12})_{LR}|, |(\delta^u_{12})_{RL}|\le 4.6 \times 10^{-3}, \qquad \text{for} \quad M_{\tilde{q}}=250\; \text{GeV}.
\end{equation}
The bounds on mass insertions $(\delta_{12}^u)_{LL}$ and $(\delta_{12}^u)_{RR}$
can be obtained from the experimental upper bound on the mass difference in the neutral D system. Saturating the experimental bound $\Delta m_D<4.5 \times 10^{-14}$ GeV \cite{Godang:1999yd,Link:2000cu} by the gluino exchange gives \cite{Fajfer:2001sa,Prelovsek:2000xy,Gabbiani:1996hi}
\begin{equation}
|(\delta^u_{12})_{LL}|\le 0.03, \qquad \text{for} \quad M_{\tilde{g}}=M_{\tilde{q}}=250\; \text{GeV}, 
\end{equation}
where $(\delta^u_{12})_{RR}$ has been set to zero. These translate into
 \begin{subequations}
\begin{align}
|V_{cb}^* V_{ub} \delta C_7| & \le 0.04, &|V_{cb}^* V_{ub} \delta C_7'| & \le 0.04, \label{c7bound}\\
|V_{cb}^* V_{ub} \delta C_9|&\le 0.0016, 
 &|V_{cb}^* V_{ub} \delta C_9'|&\simeq 0.
\end{align}
\end{subequations}
Note that both $C_7$ and $C_7'$ receive largest contributions from $(\delta^u_{12})_{LR}$ insertions. Note also that the upper limits on $C_{7}$ coefficient is three orders of magnitude larger than the Standard Model value, while for $C_9$ is an order of magnitude larger than the SM value. However, as discussed in previous section, SM prediction  is dominated by $Q_{1,2}$ insertions and therefore by long distance effects. 

 The  contributing diagrams  are shown on Fig. \ref{SDdiagr}, to which the diagrams with $Q_i\to Q_i'$ should be added. In mass insertion approximation the coefficient $C_{10}$ is small  and will be  neglected in the following.  When $Q_{7,9}$ operators are inserted, the photon  bremsstrahlung off the final lepton pair is not possible. In the case of $Q_7$ operator this is because the diagrams are of the type shown in Fig. \ref{TwoBlobs}, while in the case of $Q_9$ operator  the bremsstrahlung is prohibited because of  vector  current conservation.

Taking the values of induced Wilson coefficients at the upper bounds we obtain 
\begin{equation}
\text{Br}(D^0\to e^+e^-\gamma)_{\text{MSSM}}=1.4 \times 10^{-9}, \qquad \text{Br}(D^0\to \mu^+\mu^-\gamma)_{\text{MSSM}}=1.2 \times 10^{-9}.  \label{MSSM-pred}
\end{equation}
The  MSSM contribution to the decay rate is entirely due to  gluino exchange enhancement of $C_7,C_7'$ coefficients. The decay rate is thus enhanced in low $p^2$ region, which also explains larger increase of $D^0\to e^+e^- \gamma$ decay rate. The increase is, however, not significant enough to dominate over the resonant contributions \eqref{reson1}-\eqref{reson3}. MSSM effects, if any, are thus too small to be unambiguously detected
experimentally in the decays $D^0\to l^+l^- \gamma$. 

\subsection{R parity violation}
The situation is  quite different once the assumption of $R$ parity conservation is relaxed and the soft symmetry breaking terms are introduced. We follow the analysis of Ref. \cite{Burdman:2001tf}. The tree level exchange of down squarks results in the effective interaction
\begin{equation}
{\cal L}_{\text{eff}}=\frac{\tilde{\lambda}_{i2k}' \tilde{\lambda}_{i1k}'}{2 M_{\tilde{d}_R^k}^2} (\bar{u}_L \gamma^\mu c_L) (\bar{l}_L \gamma^\mu l_L), \label{R-inter}
\end{equation}
where $\tilde{\lambda}_{ijk}'$ are the coefficients of lepton--up-quark--down-squark $R$ parity breaking terms of the superpotential in the quark mass basis. The effective interaction \eqref{R-inter} translates into the additional contributions $\delta C_i$ to $C_{9,10}$ Wilson coefficients
\begin{equation}
V_{cb}^* V_{ub} \delta  C_9=-V_{cb}^* V_{ub} \delta  C_{10}=\frac{2 \sin^2 \theta_W}{\alpha_{\text{QED}}^2}\left(\frac{m_W}{M_{\tilde{d}_R^k}}\right)^2 \tilde{\lambda}_{i2k}' \tilde{\lambda}_{i1k}',
\end{equation}
while no contributions are generated to $C_{9,10}'$ Wilson coefficients  \cite{Burdman:2001tf}. For electrons in the final states we use bounds on $\tilde{\lambda}_{i2k}', \tilde{\lambda}_{i1k}'$ from charged current universality \cite{Allanach:1999ic}
\begin{equation}
\tilde{\lambda}_{11k}'\le 0.02 \left(\frac{M_{\tilde{d}_R^k}}{100 \text{ GeV}}\right),\qquad \tilde{\lambda}_{12k}'\le 0.04 \left(\frac{M_{\tilde{d}_R^k}}{100 \text{ GeV}}\right).\label{el-tri}
\end{equation}
 For muons in the final state, the limits come from $D^+\to \pi^+\mu^+\mu^-$ \cite{Burdman:2001tf}. Using the new experimental  bound 
$\text{Br}(D^+\to \pi^+\mu^+\mu^-)<8.8 \times 10^{-6}$ \cite{Johns:2002hd}, this gives
\begin{equation}
\tilde{\lambda}_{22k}', \tilde{\lambda}_{21k}'\le 0.003 \left(\frac{M_{\tilde{d}_R^k}}{100 \text{ GeV}}\right)^2.\label{mu-tri}
\end{equation}
The bounds on trilinear couplings \eqref{el-tri},\eqref{mu-tri} then give the following bounds on possible enhancements of $C_{9,10}$ Wilson coefficients for the electron or muon channel
\begin{subequations}\label{bounds}
\begin{align}
|V_{cb}^* V_{ub} &\delta  C_{9,10}^e|\le 4.4,\\
|V_{cb}^* V_{ub} &\delta  C_{9,10}^\mu|\le 17,
\end{align}
\end{subequations}
with $\delta C_9^{e,\mu}=-\delta C_{10}^{e,\mu}$. Note that in \eqref{bounds} the squark mass cancels. These are then added to the Standard Model values. The diagrams are listed on Figure \ref{SDdiagr}.  The possible enhancement over SM branching ratio predictions is quite striking and is in the case of muons in the final state by almost two orders of magnitude, if the values  of $C_{9,10}$ Wilson coefficients are taken to be the upper bounds in \eqref{bounds}. The diagrams on Fig. \ref{SDdiagr} with  photon bremsstrahlung off the final lepton pair and the insertion of the $Q_{10}$ operator are IR divergent. We take  cutoff energy  to be $E_{\gamma}\ge 50$ MeV or $E_{\gamma}\ge 100$ MeV . The contributions from various sources, the nonresonant \eqref{nonres-res} and resonant \eqref{reson1}-\eqref{reson3} SM contributions, the insertion of $Q_7,Q_7'$ operator with the $C_7,C_7'$ values given in \eqref{c7bound}, and the contributions  from insertion of $Q_{9,10}$ operators with $C_{9,10}^{e,\mu}$ bounded by \eqref{bounds}  are summarized in
Table \ref{MSSM-res}.

The maximal branching ratios obtainable in the framework of MSSM with $R$ parity violation are
\begin{align}
\text{Br}(D^0\to e^+e^-\gamma)_{E_{\gamma}\ge 50\; \text{MeV}}^{\not R}&=4.5 \times 10^{-9}, \qquad& \text{Br}(D^0\to \mu^+\mu^-\gamma)_{E_{\gamma}\ge 50\; \text{MeV}}^{\not R}&=50  \times 10^{-9},\\
\text{Br}(D^0\to e^+e^-\gamma)_{E_{\gamma}\ge 100\; \text{MeV}}^{\not R}&=4.5\times 10^{-9}, \qquad &\text{Br}(D^0\to \mu^+\mu^-\gamma)_{E_{\gamma}\ge 100\; \text{MeV}}^{\not R}&=46  \times 10^{-9}.
\end{align}
These are to be compared with the SM predictions \eqref{SM-pred}. Note that the SM predictions are not affected by the cuts on the soft photon energy at the order of $E_{\gamma}\ge 100\; \text{MeV}$, as the bulk of contribution either comes from the resonances or the low $p^2$ region (while the cut on $E_\gamma$ is the cut on the high $p^2$ region).

\begin{table}[h]
\begin{center}
\begin{tabular}{|c|c|c|} \hline
Contrib. & $\text{Br}(D^0\to e^+e^-\gamma)$ & $\text{Br}(D^0\to \mu^+\mu^-\gamma)$\\ \hline\hline
Nonres. & $12.9 \times 10^{-11}$ & $2.1 \times 10^{-11}$\\ \hline
Reson. &$1.1\times 10^{-9}$ &$1.1\times 10^{-9}$ \\ \hline
$C_7$ &$0.23\times 10^{-9}$  & $0.04\times 10^{-9}$ \\ \hline
$C_9$ &$1.37 \times 10^{-9}$  &$20.5\times 10^{-9}$  \\ \hline
$C_{10}$ & $1.37 \times 10^{-9}$ &   $31.3 \times 10^{-9}$\\ \hline\hline
All & $4.52 \times 10^{-9}$ &$50.2  \times 10^{-9}$  \\ \hline
\end{tabular}
\caption{\footnotesize{ The relative sizes of various possible contributions in the context of MSSM with $R$ parity violation. The photon energy cutoff is taken to be  $E_{\gamma}\ge 50$ MeV. Largest possible effects are calculated. The values for nonresonant (Nonres.) and resonant (Reson.) LD contributions are the same as for the  SM prediction. The $C_7$ denote $Q_7, Q_7'$, while $C_{9,10}$ denote $Q_{9,10}$ insertions respectively. In the last row the maximal calculated branching
ratios are given.}}\label{MSSM-res}
\end{center}
\end{table}

The enhancement due to possible $R$ parity violating contributions  is by more than an  order of magnitude in the muon channel compared to the SM prediction. The enhancement also has a distinct signal in the $d\Gamma/dp^2$ decay width distribution. In the SM model the decay $D^0\to l^+l^-\gamma$ either proceeds through $\rho, \omega,\phi$ vector resonances or through nonresonant two-meson exchanges, which are important in the low $p^2$ region. The $R$ parity violating signal on the other hand would arise from insertion of $Q_{9,10}$ operators and  is large in the region of high $p^2$ (small photon energy) region as can be seen from Fig. \ref{fig-res}.  The largest possible effect, however,  is below   expected experimental sensitivities for  rare charm decays at B-factories and CLEO-c, which are apparently  expected to be of the order of $10^{-6}$.

\section{Summary}
\label{Summary}
In this paper we have presented a detailed study of $D^0\to e^+ e^- \gamma$ and $D^0\to \mu^+\mu^- \gamma$ decays both in the Standard Model (SM) and in the Minimal Supersymmetric Standard Model (MSSM) with and without $R$ parity violation. For the SM prediction we have carried out a
calculation of the RGE  improved Wilson coefficients of $c\to u$  penguin operators, where $C_9(m_c)$ has been calculated for the first time. The penguin operators are suppressed by $V_{cb}^* V_{ub}\sim 10^{-4}$ CKM matrix elements and  are therefore irrelevant for the processes considered. The decays are dominated by the inclusion of $Q_{1,2}$ operators, which induce nonperturbative long distance (LD) effects. Nonresonant LD contributions are evaluated by employing the combined heavy quark and chiral symmetries. They are found to be important only in the region of low final lepton pair mass, while their contribution to the integrated decay width is of about 10\%, or even less for the muonic
channel.  The decay width is dominated by the cascade  decay  $D^0\to V\gamma\to l^+l^- \gamma$, where $V=\rho, \omega, \phi$. The Standard Model branching ratio is then predicted to be
\begin{equation}
\text{Br}(D^0\to l^+l^- \gamma)_{\text{SM}}=(1-3)\times 10^{-9}.
\end{equation}
We also investigated possible enhancements of the decay widths due to new physics contributions. We have found that possible effects coming from gluino-squark exchanges in the context of MSSM with $R$ parity conserved are masked by the LD contributions from SM. However, if the assumption of $R$ parity conservation is relaxed,  the tree level exchange of down squarks can increase the predicted branching ratios by more than an order of magnitude. The largest possible effect comes from the diagrams with photon bremsstrahlung off the leptons in the final state and is IR divergent. Choosing two different cuts on the photon energy we arrive at 
\begin{align}
\text{Br}(D^0\to e^+e^-\gamma)_{E_{\gamma}\ge 50\; \text{MeV}}^{\not R}&=4.5 \times 10^{-9}, \qquad& \text{Br}(D^0\to \mu^+\mu^-\gamma)_{E_{\gamma}\ge 50\; \text{MeV}}^{\not R}&=50 \times 10^{-9},\\
\text{Br}(D^0\to e^+e^-\gamma)_{E_{\gamma}\ge 100\; \text{MeV}}^{\not R}&=4.5 \times 10^{-9}, \qquad &\text{Br}(D^0\to \mu^+\mu^-\gamma)_{E_{\gamma}\ge 100\; \text{MeV}}^{\not R}&=46 \times 10^{-9}.
\end{align}
Allowing for the uncertainty in the SM calculation which we discussed
after Eq.\eqref{SM-pred}, we consider that branching ratios in excess of $0.5\times 10^{-8}$
are not accountable by the SM. The effect of MSSM with R parity violation
in the muon channel is the closest to the experimental sensitivities
expected at B-factories and CLEO-c. Thus we propose the $D^0\to \mu^+\mu^-\gamma$ 
decay as a possible probe of new physics.

\section*{Acknowledgments}
The research of S.F. and J.Z. was supported in part by the Ministry of
Education, Science and Sport of the Republic of Slovenia. The research
of P.S. was supported in part by Fund for Promotion of Research at the
Technion. P.S. is grateful to the High Energy Physics Group at
University College London for the hospitality during summer 2002,
when this work was completed.

\appendix
%
%
\setcounter{equation}{0}%
\renewcommand{\theequation}{\mbox{\Alph{section}.\arabic{equation}}}%

\section{Calculation of Wilson coefficients}\label{app-Wilson}
In this section we outline the calculation of Wilson coefficients listed in Table \ref{tab-Wilson}. In general, the Wilson coefficients at lower scale are calculated through the following steps. First the Wilson coefficients $C_i(m_W)$ at weak scale are calculated by matching the effective theory with five active flavors $q=u,d,s,c,b$ onto the full theory. Then the anomalous dimensions $\gamma^{(5)}$ are calculated in the effective theory with five flavors. Using $\gamma^{(5)}$, the Wilson coefficients are evolved down using RGE to the  $b$-quark scale, obtaining  $C_i(m_b)$. Then   $b$-quark is integrated out as an effective degree of freedom. This is accomplished by matching the effective theory with five flavors onto effective theory with four flavors. Remaining Wilson coefficients are then evolved down to charm scale using anomalous dimension matrices of four-flavor effective theory. Thus
\begin{equation}
\vec{C}(m_c)=U_4(m_c,m_b)M_5(m_b) U_5(m_b,m_W)\vec{C}(m_W),
\end{equation}
where $U_{4,5}(\mu_1,\mu_2)$ are evolution matrices from scale $\mu_2$ to scale $\mu_1$ in four and five-flavor effective theories respectively, while $M_5$ is the threshold matrix that matches the two effective theories at scale $\mu\sim m_b$.

As already discussed in section \ref{Eff-weak-lagr},  $Q_{10}$ does not mix with other operators due to chirality, so that $C_{10}(\mu_c)=C_{10}(\mu_W)$. Also, the dimension five operators $Q_{7,8}$ do not mix into dimension six operators $Q_{1,\dots,6}$ and $Q_{9}$. If one is interested in these operators solely, the dimension five operators can be dropped from the RG analysis. We will follow this procedure and evaluate $C_7$ separately. Note also, that (i) $Q_9$ operator does not mix into operators $Q_{1,\dots,6}$ and (ii) penguin operators $Q_{3,\dots,6}$ do not mix into operators $Q_{1,2}$. One can thus consider the RG evolution of reduced operator basis $Q_{1,2}$, $Q_{1, \dots,6}$ or $Q_{1,\dots,6,9}$ if one is interested in smaller sets of Wilson coefficients $C_{1,2}$, $C_{1,\dots,6}$, or $C_{1,\dots,6,9}$ without introducing any error in the calculation. Finally, it is convenient to introduce a rescaled operator $\tilde{Q}_9=\alpha/\alpha_s (\bar{u}c)_{V-A} (\bar{l}l)_V$ \cite{Buchalla:1995vs}, as then the anomalous dimension depend only on strong coupling.

It is instructive to do the $\alpha_s$ counting. At leading order the RG evolution sums terms of form $\alpha_s \ln(m_c^2/\mu_W^2)$ which are numerically of order ${\cal O}(1)$. At leading order one thus has to start with initial values $C_i(m_W)$ calculated at $\alpha_s^0$, and then evolve them using 1 loop anomalous dimensions (i.e. of order $\alpha_s$) to get order ${\cal O}(1)$ values $C_i(\mu)$ at lower scales. Going to higher orders an additional power of $\alpha_s$ is added at each step. We thus have
\begin{equation}
C_i(\mu)={\cal O}(1)+{\cal O}(\alpha_s)+\dots
\end{equation}
This expansion is valid also for $\tilde{C}_9$ multiplying the rescaled operator $\tilde{Q}_9$. Since $Q_9=\alpha_s/(8\pi) \tilde{Q}_9$, then $C_9=8 \pi/\alpha_s \tilde{C}_9$, so that the expansion is
\begin{equation}
C_9(\mu)={\cal O}(1/\alpha_s)+{\cal O}(1)+{\cal O}(\alpha_s)+\dots \label{C9exp}
\end{equation}
It is thus only the NLO term that is of order ${\cal O}(1)$ in the calculation of $C_9$ Wilson coefficient. It is then consistent in $\alpha_s$ counting to work with $C_9$ determined at NNLO and with other Wilson coefficients at NLO (if one wishes to work to ${\cal O}(\alpha_s)$). Partial calculations at NNLO became available in the literature recently \cite{Asatrian:2001de,Asatryan:2001zw,Bobeth:1999mk,Lunghi:2002qw,Ghinculov:2002pe}, however, the three-loop calculation of NNLO dimensional matrix has still not been performed. For this reason we will work in the following with both $C_9$ and $C_{1,\dots,6}$ determined at NLO.

The effective Lagrangian at weak scale $\mu\sim m_W$ is
\begin{equation}
\begin{split}
{\cal L}_{\text{eff}}=-\frac{G_F}{\sqrt{2}} \big[ &V_{cd}^* V_{ud} \big(\sum_{i=1,2}C_i Q_i^d+\sum_{i=3,\dots,6,9}C_i Q_i\big)+\\
+&V_{cs}^* V_{us} \big(\sum_{i=1,2}C_i Q_i^s+\sum_{i=3,\dots,6,9} C_iQ_i\big)+\\
+&V_{cb}^* V_{ub} \big(\sum_{i=1,2}C_i Q_i^b+\sum_{i=3,\dots,6,9}C_i Q_i\big)\big]\\
=-\frac{G_F}{\sqrt{2}} \big[ &V_{cd}^* V_{ud} \sum_{i=1,2}C_i \big(Q_i^d-Q_i^b\big)
+V_{cs}^* V_{us} \sum_{i=1,2}C_i \big(Q_i^s-Q_i^b\big)\big], \label{lagr-at-weak}
\end{split}
\end{equation}
where $Q_{1,2}^{d,s,b}$ and the penguin and electromagnetic operators $Q_i$, $i=3,\dots,6,9$ are defined in Eqs. \eqref{list-oper}. The contributions of electromagnetic penguins have been neglected as they are suppressed by additional powers of $\alpha$ in the processes considered. In the last line of \eqref{lagr-at-weak} the unitarity of CKM matrix has been used $V_{cd}^* V_{ud}+V_{cs}^* V_{us}+V_{cb}^* V_{ub}=0$. Above, also the masses of $d,s,b$ quarks have been neglected compared to the weak scale, so that the Wilson coefficients $C_i$ are the same regardless of the flavor of down quark flowing in the loop in the full theory (i.e. regardless of the CKM structure in front of the parenthesis in \eqref{lagr-at-weak}). Thus the penguin operators do not appear in the effective Lagrangian at the weak scale as long as the mass of $b$-quark can be neglected compared to $m_W$. This is in contrast to the case of $\Delta B=1$ decays, where up-type quarks flow in the loops in the full theory. Since the top quark is very heavy, its mass cannot be neglected in the loops. This induces penguin operators already at the weak scale.

We  start with the values of Wilson coefficients to order $\alpha_s$ at weak scale. These are known for quite some time and are (in naive dimensional regularization scheme (NDR) \cite{Buras:1989xd})
\begin{align}
C_1(m_W)&=\frac{11}{2} \frac{\alpha_s(m_W)}{4 \pi}, &C_2(m_W)&=1-\frac{11}{6}\frac{\alpha_s(m_W)}{4 \pi},
\end{align}
while $C_{3,\dots,9}(m_W)=0$. Above $\mu_b$ the penguin operators do not enter the effective Lagrangian due to unitarity of CKM matrix.  The Wilson coefficients $C_{1,2}$  are  evolved down to $\mu\sim \mu_b$ using $2\times  2$ anomalous dimension matrix (that can be found in \cite{Buras:1989xd} or in Eq. (5.12) of \cite{Buchalla:1995vs}).  At the scale $\mu_b$  the $b$-quark is integrated out, i.e. the five-flavor effective theory \eqref{lagr-at-weak} is matched onto the four-flavor theory with the Lagrangian
\begin{equation}
\begin{split}
{\cal L}_{\text{eff}}=-\frac{G_F}{\sqrt{2}} \big[ &V_{cd}^* V_{ud} \big(\sum_{i=1,2}C_i Q_i^d+\sum_{i=3,\dots,6,9}C_i Q_i\big)+ V_{cs}^* V_{us} \big(\sum_{i=1,2}C_i Q_i^s+\sum_{i=3,\dots,6,9} C_iQ_i\big)\big]\\
=-\frac{G_F}{\sqrt{2}} \big[ &V_{cd}^* V_{ud} \sum_{i=1,2}C_i Q_i^d+ V_{cs}^* V_{us} \sum_{i=1,2}C_i Q_i^s-V_{cb}^* V_{ub} \sum_{i=3,\dots,6,9} C_iQ_i\big]. \label{lagr-at-bandc}
\end{split}
\end{equation}
The penguin operator Wilson coefficients $C_{3,\dots,6,9}$ arise from the matching procedure. This is the only nontrivial step in the application of formulas from the literature as these were calculated for down-quark transitions. We use expressions for $K_L\to \pi^0 e^+ e^-$ decay \cite{Buras:qa} where a similar matching procedure has to be done at charm mass, with the c-quark being integrated out. The results of Ref. \cite{Buras:qa} apply directly for the matching of gluonic penguin operators also in the case considered here, when $b$-quark is integrated out, while the semileptonic Wilson coefficient $C_9$ has to be multiplied by $e_b/e_c=-1/3 \cdot 3/2=-1/2$. We then have (see Eqs. (6.20), (8.9) of \cite{Buchalla:1995vs})
\begin{align}
Z_1(m_b)&=C_1(m_b), &Z_2(m_b)&=C_2(m_b),\label{Z-begin}\\
Z_3(m_b)&=-\frac{\alpha_s}{24 \pi} F_s(m_b), &Z_4(m_b)&=\frac{\alpha_s}{8 \pi} F_s(m_b),\\
Z_5(m_b)&=-\frac{\alpha_s}{24 \pi} F_s(m_b), &Z_6(m_b)&=\frac{\alpha_s}{8 \pi} F_s(m_b),\\
Z_9(m_b)&=-\frac{1}{2} Z_{7V}'(m_b)=\frac{\alpha_s}{4 \pi} F_e(m_b), &{}&\label{Z-end}
\end{align}
with $Z_{7V}'$ defined as in Eq. (8.9) of \cite{Buchalla:1995vs}, while the functions
\begin{align}
F_s(\mu)&=-\frac{2}{3} \Big[\ln\Big(\frac{m_b^2}{\mu^2}\Big)+1\Big]Z_2(\mu),\\
F_e(\mu)&=-\frac{4}{9} \Big[\ln\Big(\frac{m_b^2}{\mu^2}\Big)+1\Big]\big(3Z_1(\mu)+ Z_2(\mu)\big),
\end{align}
are again calculated in NDR (see also Eqs. (4.29)-(4.31) of \cite{Buras:1993dy}).

 The sets of operators $\{ Q_{1,2}^d, Q_{3,\dots,6}, Q_9\}$ and $\{ Q_{1,2}^s, Q_{3,\dots,6}, Q_9\}$  from the first line  of \eqref{lagr-at-bandc} are then evolved to the charm scale $\mu\sim m_c$ using $7\times  7$ anomalous dimension  matrices $\gamma$ for four quark effective theory. The $6\times  6$ LO and NLO submatrices involving gluonic penguins are listed in Eqs. (6.25), (6.26) of Ref. \cite{Buchalla:1995vs} and have been calculated in \cite{Buras:1991jm,Ciuchini:1993vr}. The remaining entries are listed in Eqs (8.11), (8.12) of Ref. \cite{Buchalla:1995vs} and have been calculated in \cite{Buras:qa}. 

In summary, the RG evolution from $\mu_W$ to $\mu_c$ for $\Delta C=1$ transitions is described by the following procedure
\begin{align}
m_b< \mu< m_W:& \qquad  \vec{C}(\mu)=U_5(\mu,m_W) \vec{C}(m_W),\label{RG-flow-beg}\\
\mu=m_b:&\qquad \vec{C}(m_b)\to \vec{Z}(m_b),\\
m_c< \mu< m_b:& \qquad  \vec{C}(\mu)=U_4(\mu,m_b) \vec{Z}(m_b),\label{RG-flow-end}
\end{align}
with $U_5$ and $U_4$ the $2\times  2$  and $7\times  7$ evolution matrices for five and four active flavors respectively. They can be found in Eqs. (3.93)-(3.98) of \cite{Buchalla:1995vs}. The  $Z(m_b)$ are given in \eqref{Z-begin}-\eqref{Z-end}. The values of calculated Wilson coefficients are listed in Table \ref{tab-Wilson}.

We next turn to the calculation of effective parameters $C_{7,9,10}^{\text{IL}}$ corresponding to invariant amplitudes calculated using full theory but neglecting QCD interactions. Consider $c\to u\gamma$ and $c\to u l^+l^-$ invariant amplitudes corresponding to the diagrams on Fig. \ref{InamiLim}. The invariant amplitudes obtained neglecting QCD would have the same structure as one would get from the operators $Q_{7,9}$ in the effective Lagrangian \eqref{weak-eff-lagr}, if used {\it at tree level}. The parameters corresponding to these  invariant amplitudes will be denoted $C_{7,9}^{\text{IL}}$. It is important to stress that these are {\it not Wilson coefficients}, as they only parametrize invariant amplitudes.  They are easily obtained  using calculation of Ref. \cite{Inami:1980fz} for $b\to s l^+l^-$ transitions. Following \cite{Ho-Kim:1999bs} we find that the coefficients are of the form
\begin{equation}
C_n= I_l F_I(x_i)+Q_l F_Q(x_i),\label{Cnmass-dep}
\end{equation}
where $I_l$ is connected to the weak isospin of the quarks in the loops of Fig. \ref{InamiLim}. and $Q_l$ is their charge. For up quarks in the loops, as is the case for $b\to sl^+l^-$ we have $I_l=+1$, $Q_l=2/3$. For the case $c\to u l^+l^-$, that we are interested in, $I_l=-1$, $Q_l=-1/3$, as then down quarks appear as intermediate states in the loops. $F_I(x_i)$ and $F_Q(x_i)$ are functions of CKM matrix elements and masses of quarks running in the loops, $x_i=m_{q_i}^2/m_W^2$. The functions $F_I(x_i)$ and $F_Q(x_i)$ have been determined by Inami and Lim  \cite{Inami:1980fz}. Using their definitions one arrives at
\begin{align}
C_9^{\text{IL}}&=-\frac{\tilde{C}}{2}\frac{1}{\sin^2\theta_W} - \frac{\tilde{H}_1}{4},\\
C_{10}^{\text{IL}}&=\frac{\tilde{C}}{2} \frac{1}{\sin^2\theta_W},
\end{align}
with
\begin{equation}
\tilde{C}=-4 \sum_{j=s,b} \lambda_j \bar{C}(x_j,x_d)I_l, \qquad
\tilde{H}_1= 16\sum_{j=s,b}\lambda_j \big[ \bar{F}_1(x_j,x_d)+2 \bar{\Gamma}_z(x_j,x_d) I_l\big],
\end{equation}
where $\lambda_j=V_{cj}^* V_{uj}/(V_{cb}^* V_{ub})$, the function $\bar{C}(x_j, x_d)$ is defined in Eq. (2.14) of Ref. \cite{Inami:1980fz}, the function $\bar{\Gamma}_z(x_j, x_d)$ in  Eq. (2.7) of Ref. \cite{Inami:1980fz}, while $\bar{F}_1(x_j, x_d)$ is defined in  Eq. (B.2) of Ref. \cite{Inami:1980fz} (note also the errata),  the later function being changed slightly, as now
\begin{equation}
\bar{F}_1=Q_l \big\{\dots\big\}+I_l \dots
\end{equation}
(i.e. the last two lines of Eq. (B.2) in  Ref. \cite{Inami:1980fz} are to be multiplied with $I_l$).

The important thing to note is that the variable $x_j=m_{q_j}^2/m_W^2$ is very small for $q_j=d,s,b$. The functions $\bar{C}$ and $\bar{\Gamma}_z$ are proportional to $\bar{C},\bar{\Gamma}_z\propto x_j $ and are thus very small. The function $\bar{F_1}$ on the other hand is to the leading order $\bar{F}_1(x_j, x_d)\sim \frac{2}{3} Q_l \ln(x_j/x_d)$ which is of order ${\cal O}(1)$.  Following the same procedure also the value of $C_7^{\text{IL}}$ can be obtained. The leading order  expressions in terms of  $x_j=m_{q_j}^2/m_W^2$ are then
\begin{subequations}\label{C7-c10IL}
\begin{align}
C_7^{\text{IL}}&\simeq -5/24 \sum_j \lambda_j x_j,\\
C_9^{\text{IL}}&\simeq -\lambda_s 16/9 \ln \big(m_s/m_d),\\
C_{10}^{\text{IL}}&\simeq 2\sum_j \lambda_j x_j \frac{1}{\sin^2 \theta_W}.
\end{align}
\end{subequations}

The comparison of $C_9^{\text{IL}}$ coefficient and the Wilson coefficient $C_9(m_c)$ has already been made in section \ref{Eff-weak-lagr}, where it was found that  $C_9(m_c)\sim 10^{-4}C_9^{\text{IL}} $. The situation is somewhat different in the case of $C_7$ Wilson coefficient. Using $|V_{cb}^* V_{ub}|=(1.3 \pm 0.4)\times 10^{-4}$ one gets  $|C_7^{\text{IL}}|\sim 10^{-3}$, which is an order of magnitude smaller than the RG improved Wilson coefficient $C_7$.  Namely, the RG evolution lifts  the hard GIM mechanism of $C_7\sim \sum_j \lambda_j x_j $ and  replaces it  with logarithmic dependence on the scales $\mu\sim m_c, m_b, m_W$ involved in the RG evolution. However, the $V_{cb}^* V_{ub}$ suppression of $Q_7$ operator is still present. Again, both  the inclusive rate $c\to u\gamma$ as well as exclusive decays are  dominated by the inclusion of operators $Q_{1,2}$ at one-loop level \cite{Greub:1996wn}.

Finally, using Wolfenstein CKM parameters $\rho=0.4, \eta=0.45, A=0.83$ and the quark masses $m_d=6$ MeV, $m_s=130$ MeV, $m_b=4.25$ GeV, we arrive at $C_{10}^{\text{IL}}=(3.9+1.7 i)\times 10^{-2}$. Note that (i) if masses of $d,s,b$ quarks can be neglected compared to the $m_W$, then $C_{10}=0$ and that then (ii) the low energy QCD and QED interactions cannot induce a nonzero value of $C_{10}$ Wilson coefficient. It is thus consistent with the assumptions of OPE to set $C_{10}=0$ as has been done in this paper.

\section{List of the chiral loop integrals}\setcounter{equation}{0}%
\label{app-A}
In this appendix we list definitions of the  dimensionally regularized integrals needed in the  evaluation
of
$\chi$PT and HQ$\chi$PT one-loop graphs shown on Fig.
\ref{fig-1}. The integrals containing heavy quark propagator are
\begin{align}
 -\frac{1}{16 \pi^2} \bar{A}_0 (m&)=\frac{i \mu^{\epsilon}}{(2\pi)^{n}} \int d^{n}q \frac{1}{(v \negcdot q -\Delta+i\delta)}=0,\\
 -\frac{1}{16 \pi^2} \bar{B}_{\{0,\mu,\mu\nu\}} (m&,\Delta)=\frac{i \mu^{\epsilon}}{(2\pi)^{n}}  \int d^{n}q \frac{\{1,q_\mu,q_{\mu}q_{\nu}\}}{(v \negcdot q -\Delta+i\delta)(q^2-m^2+i \delta)},
\\
\begin{split}
-\frac{1}{16 \pi^2} \bar{C}_{\{0,\mu,\mu\nu\}} (&p, m_1,m_2, \Delta) =\\
&\frac{i \mu^\epsilon}{(2 \pi)^{n}} \int d^{n} q \frac{\qquad \qquad \qquad\quad\{1,q_\mu,q_\mu q_\nu\}\hfill}{(v \negcdot q -\Delta)(q^2-m_1^2 )((q+p)^2-m_2^2)} ,\label{CBar}
\end{split}
\\
\begin{split}
-\frac{1}{16 \pi^2} \bar{D}_{\{0,\mu,\mu\nu\}}(&p_1,p_2,m_1,m_2,m_3,\Delta)=\\
&\frac{i \mu^\epsilon}{(2\pi)^{n}} \int \frac{\; \; d^{n}q \qquad \qquad \qquad\{1,q_\mu,q_\mu q_\nu\} \hfill }{(v\negcdot q-
\Delta)(q^2-m_1^2)((q+p_1)^2-m_2^2)((q+p_2)^2-m_3^2)},\label{DBar}
\end{split}
\end{align}
where $n=4-\epsilon$. The dependence of scalar and tensor functions on $v^\mu$ is not shown explicitly and  also in Eqs. \eqref{CBar},\eqref{DBar} the $i\delta$ prescription is not shown.  The scalar integrals $\bar{B}_0(m,\Delta)$, $\bar{C}_{0} (p, m_1,m_2, \Delta)$, $\bar{D}_{0}(p_1,p_2,m_1,m_2,m_3,\Delta) $ have been calculated in \cite{Zupan:2002je}. We use the expressions of Ref.  \cite{Zupan:2002je} in the numerical evaluation of scalar integrals $\bar{B}_0$, $\bar{C}_0$, $\bar{D}_0$. The tensor integrals can be expressed in terms of Lorentz-covariant tensors. The notation we use for the tensor functions resembles closely the notation used in  Ref. \cite{Hahn:1998yk} for the Veltman-Passarino  functions \cite{Passarino:1978jh}  
\begin{align}
\bar{B}_\mu (m,\Delta)&= v_\mu\bar{B}_1 \label{tensor-beg},\\
 \bar{B}_{\mu\nu}(m,\Delta)&=\eta_{\mu\nu}\bar{B}_{00} +v_\mu v_\nu \bar{B}_{11},\\
\bar{C}_\mu(p,m_1,m_2,\Delta)&= v_\mu\bar{C}_1+p_\mu\bar{C}_2,\\
\bar{C}_{\mu\nu}(p,m_1,m_2,\Delta)&=\eta_{\mu\nu}\bar{C}_{00}+(v_\mu p_\nu+p_\mu v_\nu)\bar{C}_{12}+v_\mu v_\nu\bar{C}_{11}+p_\mu p_\nu \bar{C}_{22},\\
\bar{D}_\mu(p_1,p_2,m_1,m_2,m_3,\Delta)&= v_\mu\bar{D}_1+p_{1\mu}\bar{D}_2+p_{2\mu}\bar{D}_3,\\
\begin{split}
\bar{D}_{\mu\nu}(p_1,p_2,m_1,m_2,m_3\Delta)&=\eta_{\mu\nu}\bar{D}_{00}+v_\mu v_\nu\bar{D}_{11}+(v_\mu p_{1\nu}+p_{1\mu} v_\nu)\bar{D}_{12}\\
&\qquad +p_{1\mu} p_{1\nu} \bar{D}_{22}+(v_\mu p_{2\nu}+p_{2\mu} v_\nu)\bar{D}_{13}+p_{2\mu} p_{2\nu} \bar{D}_{33},\label{tensor-end}
\end{split}
\end{align}
The tensor functions are calculated using the algebraic reduction \cite{Passarino:1978jh}, i.e. the tensor functions  \eqref{tensor-beg}-\eqref{tensor-end} are multiplied by four-momenta $v^\mu, p^\mu, \dots$ or contracted using $\eta^{\mu\nu}$. Then the identities such as $v\negcdot q=v\negcdot q-\Delta+\Delta$ and/or $q\negcdot p=1/2 ( (q+p)^2-m^2-(q^2-m^2))$ are used to reduce tensor integrals to a sum of scalar integrals. The result of this procedure has been given explicitly in \cite{Eeg:2001un} for the case of two point functions $\bar{B}_{\{\mu, \mu\nu\}}$ \footnote{Note that different notation is used in Ref. \cite{Eeg:2001un}, with $\bar{B}_0(m,\Delta)=-I_2(m,\Delta)/\Delta$, $\bar{B}_1(m,\Delta)=-I_2(m,\Delta)-I_1(m)$, $\bar{B}_{00}(m,\Delta)=-\Delta J_1(m,\Delta)$, $\bar{B}_{11}(m,\Delta)=-\Delta J_2(m,\Delta)$.}. For the case of the three and four-point functions $\bar{C}_{\{\mu,\mu\nu\}}$, $\bar{D}_{\{\mu,\mu\nu\}}$ we do not write out explicitly the analytic  results of algebraic reductions as the expressions are relatively cumbersome. For instance in the case of $\bar{D}_{\mu\nu}$ the final expression involves the inverse of a $7\times  7$ matrix that corresponds to seven functions $\bar{D}_{00}\dots \bar{D}_{33}$ appearing in the expression of four-point tensor function \eqref{tensor-end}. Note as well that in this particular case there are ten possible relations between  $\bar{D}_{00}\dots \bar{D}_{33}$ and the scalar functions $\bar{B}_0$, $\bar{C}_0$, $\bar{D}_0$ that one gets from algebraic reductions (three equations from each multiplication by $v^\mu$, $p_1^\mu$, $p_2^\mu$ plus one relation from contraction by $\eta^{\mu\nu}$). Obviously not all ten equations can be linearly independent. Using different sets of seven independent equations have to lead to the same results for $\bar{D}_{00}\dots \bar{D}_{33}$ coefficient functions.  This fact  can then be used as a very useful check in the numerical implementation.

The aforementioned procedure runs into problems when implemented as it is in the calculation of $D^0\to  l^+l^-\gamma$. Namely, for $p_1=p$ and $p_2=p+k$ appearing in the calculation of  $C_0^{4\_4}$ (with $p$ the four-momentum of lepton pair and $k$ the photon momentum, see  Eqs.~\eqref{eta4_4}, \eqref{kp4_4}) only six out of ten relations following from algebraic reduction are linearly independent. This problem has been circumvented by first calculating the tensor four-point functions with prescription $k\to k+\epsilon a$, with $a$ some arbitrary four-momentum and then taking the  limit $\epsilon \to 0$ numerically. Similarly, in the calculation of  $C_0^{4\_5}$, where $p_1=k$, $p_2=k+p$, see Eqs.~\eqref{eta4_5}, \eqref{kp4_5}, the prescription $p\to p+\epsilon a$ has been used. Because  $\bar{D}_{00}\dots \bar{D}_{33}$ are continuous functions of $p_1$ and $p_2$, the outlined limiting procedure leads to an unambiguous result. This has been also checked numerically.

To make the paper self-contained we list in the following  also the notation for the Veltman-Passarino functions employed  by the LoopTools package \cite{Hahn:1998yk} that has been used for their numerical evaluation. A general integral is
\begin{equation}
-\frac{1}{16 \pi^2} T^N_{\mu_1\dots\mu_P}=\frac{i \mu^\epsilon}{(2\pi)^n}\int\frac{\; \; d^{n}q \qquad \qquad \qquad q_{\mu_1}\cdots  q_{\mu_P} \hfill }{(q^2-m_1^2)((q+p_1)^2-m_2^2)\cdots((q+p_{N-1})^2-m_{N-1}^2)},
\end{equation}
with two-point functions $T^2$ usually  denoted by letter $B$, the three-point functions $T^3$ by $C$ and the four-point functions $T^4$ by $D$. Thus e.g.  $B_0(p^2,m_1^2,m_2^2)$ and  $C_0(p_1^2,(p_1-p_2)^2, p_2^2, m_1^2,m_2^2)$ are two-point and three-point scalar functions respectively.  The decomposition of tensor integrals in terms of Lorentz-covariant tensors reads explicitly  
\begin{align}
B_\mu&=p_{1\mu}B_1,\\
B_{\mu\nu}&=\eta_{\mu\nu}B_{00}+p_{1\mu}p_{1\nu}B_{11},\\
C_{\mu}&=p_{1\mu}C_1+p_{2\mu}C_2=\sum_{i=1}^2p_{i\mu}C_i,\\
C_{\mu\nu}&=\eta_{\mu\nu}C_{00}+\sum_{i,j=1}^2 p_{i\mu} p_{j\nu}C_{ij},\\
C_{\mu\nu\rho}&=\sum_{i=1}^2(\eta_{\mu\nu}p_{i\rho}+\eta_{\nu\rho} p_{i\mu}+\eta_{\mu\rho} p_{i\nu})C_{00i}+\sum_{i,j,l=1}^2 p_{i\mu}p_{j\nu}p_{l\rho} C_{ijl}.
\end{align}
Note that the tensor-coefficient functions are totally symmetric in their indices.

\section{Nonresonant LD invariant amplitudes}\label{app-B}\setcounter{equation}{0}%
In this appendix we list the analytical results for the  diagrams shown on Fig.
\ref{fig-1}. They contribute only to the $M_0^{\mu\nu}$ part of the invariant amplitude \eqref{inv_ampl}. Since separate diagrams are not gauge invariant, a general form of an invariant amplitude corresponding to a {\it single} diagram is
\begin{align}
M_0^{i} &=M_0^{i\mu\nu}\epsilon_\mu^*(k)\frac{1}{p^2} \bar{u}(p_1)\gamma_\nu v(p_2),\\
M_0^{i\mu\nu}&=C_{0\eta}^i(p^2)\eta^{\mu\nu}-C_{0kp}^i(p^2)\frac{p^\mu k^\nu}{p\negcdot k}+ D_0^i(p^2) \epsilon^{\mu \nu \alpha \beta} k_\alpha p_\beta.
\end{align}
For a gauge invariant sum of diagrams therefore $\sum_iC_{0\eta}^i=\sum_iC_{0kp}^i$ (c.f. \eqref{inv_amp}) has to be true, which represents a very useful numerical check. 

Note that $D_0^i(p^2)$ form factors corresponding to diagrams on Fig. \ref{fig-1} are zero. The analytical expressions for $C^i_{0\eta,kp}(p^2)$ form factors are
\begin{align}
C_{0\eta}^{1\_1}& =i g K\Big\{ V_{us} V^*_{cs} \bar{B}_0(m_K,v\negcdot p+\Delta_s^*)+ V_{ud} V^*_{cd}\bar{B}_0(m_\pi,v\negcdot p+\Delta^*)\Big\},\\
C_{0kp}^{1\_1}&=\frac{k\negcdot p}{m_D^2} C_{0\eta}^{1\_1},\\
C_{0\eta}^{1\_2}& =-2 i  g K( V_{us} V^*_{cs} \bar{C}_{00}(-k,m_K,m_K,m_D+\Delta_s^*)+ V_{ud} V^*_{cd}\bar{C}_{00}(-k,m_\pi,m_\pi,m_D+\Delta^*)),\label{eta1_2}\\
\begin{split}
C_{0kp}^{1\_2}& =-2 i  g K \frac{k \negcdot p}{m_D^2}\Big\{ V_{us} V^*_{cs} \Big[\bar{C}_{00}(-k,m_K,m_K,m_D+\Delta_s^*) +(m_D-vk)\bar{C}_{12}(-k,m_K,\dots)\Big]\\
&\qquad\qquad\qquad+ V_{ud} V^*_{cd}\Big[ \quad m_K\to m_\pi, \;\Delta_s^*\to \Delta^* \quad\Big]\Big\},
\end{split}\label{kp1_2}
\\
C_{0\eta}^{2\_1+2\_2}&=0,
\\
\begin{split}
C_{0kp}^{2\_1+2\_2}&=- i g K \frac{1}{(v\negcdot k) (v\negcdot p)} \frac{k\negcdot p}{m_D^2}\Big\{ V_{us} V_{cs}^*\Big[ \big(m_K^2-\Delta_s^{*2}\big)\bar{B}_0(m_K,\Delta_s^*)+\big(m_K^2-(m_D+\Delta_s^*)^2\big)\\
&\times \bar{B}_0(m_K,m_D+\Delta_s^*)-\big(m_K^2-(vk+\Delta_s^*)^2)\big)\bar{B}_0(m_K,vk+\Delta_s^*)\\
&-(m_K^2-(vp+\Delta_s^*)^2)\bar{B}_0(m_K,vp+\Delta_s^*)\Big] + V_{ud} V_{cd}^* \Big[ \quad m_K\to m_\pi, \;\Delta_s^*\to \Delta^* \quad \Big]\Big\},
\end{split}
\\
C_{0\eta}^{2\_3}&=0,\\
\begin{split}
C_{0kp}^{2\_3}&= 2 ig  K \frac{k \negcdot p}{m_D^2}\Big\{
 V_{us} V_{cs}^* \frac{1}{2(m_D-vk)}\Big[ \bar{B}_1(m_K,vk+\Delta_s^*)-\bar{B}_1(m_K,m_D+\Delta_s^*)\\
&+\bar{B}_1(m_K,\Delta_s^*)-\bar{B}_1(m_K,m_D-vk+\Delta_s^*)\\
&+2\big(m_K^2-\frac{1}{2}k^2-\Delta_s^*(vk+\Delta_s^*)\big)\bar{C}_1(-k,m_K,m_K,vk+\Delta_s^*)\\
&-2\big(m_K^2-\frac{1}{2}k^2-(m_D-vk +\Delta_s^*)(m_D+\Delta_s^*)\big)\bar{C}_1(-k,m_K,m_K,m_D+\Delta_s^*)\Big]
\\
& + V_{ud} V_{cd}^* \frac{1}{2(m_D-vk)}\Big[ \quad m_K\to m_\pi, \;\Delta_s^*\to \Delta^* \quad \Big]\Big\},
\end{split}
\\
C_{0\eta}^{3\_1}&=C_{0\eta}^{1\_1}( \rm{with}\; k \leftrightarrow p),\\
C_{0kp}^{3\_1}&=C_{0kp}^{1\_1}( \rm{with}\;  k \leftrightarrow p),
\end{align}
\begin{align}
C_{0\eta}^{3\_2}&=-2i g K \Big[V_{us} V_{cs}^* \bar{C}_{00}(k,m_K,m_K,\Delta_s^*)+V_{ud} V_{cd}^* \bar{C}_{00}(k,m_\pi,m_\pi,\Delta^*)\Big],\label{eta3_2}\\
\begin{split}
C_{0kp}^{3\_2}&=-2i g K\frac{k\negcdot p}{m_D^2}\Big[V_{us} V_{cs}^*\big( \bar{C}_{00}(k,m_K,m_K,\Delta_s^*)-(m_D-vk)\bar{C}_{12}(k,m_K,m_K,\Delta_s^*)\big)\\
&\qquad\qquad\quad+V_{ud} V_{cd}^* \big(\quad m_K\to m_\pi, \;\Delta_s^*\to \Delta^* \quad\big)\Big],
\end{split}\label{kp3_2}
\\
C_{0\eta}^{4\_1}&=C_{0\eta}^{1\_2}( \rm{with}\; k \leftrightarrow p),\\
C_{0kp}^{4\_1}&=C_{0kp}^{1\_2}( \rm{with}\;  k \leftrightarrow p),\\
C_{0\eta}^{4\_2}&=C_{0\eta}^{2\_3}( \rm{with}\; k \leftrightarrow p),\\
C_{0kp}^{4\_2}&=C_{0kp}^{2\_3}( \rm{with}\;  k \leftrightarrow p),
\end{align}
\begin{align}
C_{0\eta}^{4\_3}&=C_{0\eta}^{3\_2}( \rm{with}\; k \leftrightarrow p),\\
C_{0kp}^{4\_3}&=C_{0kp}^{3\_2}( \rm{with}\;  k \leftrightarrow p),
\\
C_{0\eta}^{4\_4}&=4 i g K\Big[V_{us} V_{cs}^* f_\eta (p,k,m_K,\Delta_s^*)+V_{ud} V_{cd}^* f_\eta(p,k,m_\pi,\Delta^*)\Big],\label{eta4_4}\\
C_{0kp}^{4\_4}&=-4 i g K \frac{k\negcdot p}{m_D^2}\Big[V_{us} V_{cs}^* f_{kp} (p,k,m_K,\Delta_s^*)+V_{ud} V_{cd}^* f_{kp}(p,k,m_\pi,\Delta^*)\Big],\label{kp4_4}\\
C_{0\eta}^{4\_5}&=C_{0\eta}^{4\_4}( \rm{with}\; k \leftrightarrow p),\label{eta4_5}\\
C_{0kp}^{4\_5}&=C_{0kp}^{4\_4}( \rm{with}\;  k \leftrightarrow p),\label{kp4_5}
\\
\begin{split}
C_{0\eta}^{4\_6}&=2 i g K\Big\{V_{us} V_{cs}^*\Big[- \bar{B}_0(m_K,m_D+\Delta_s^*)-\bar{C}_0(p+k,m_K,m_K,\Delta_s^*)\times\\
&\times(m_K^2-\Delta_s^{*2}) +m_D B_1(m_D^2,m_K^2,m_K^2)+\Delta_s^*B_0(m_D^2,m_K^2,m_K^2)\Big]
\\
&\qquad + V_{ud} V_{cd}^* \Big[ \quad m_K\to m_\pi, \;\Delta_s^*\to \Delta^* \quad \Big]\Big\},
\end{split}
\\
C_{0kp}^{4\_6}&=0,
\\
\begin{split}
C_{0\eta}^{5\_1+5\_2}&= 2 i K m_D \Big\{ V_{us}V_{cs}^* \Big[\frac{m_K^2}{2}C_0(0,p^2,m_D^2,m_K^2,m_K^2,m_K^2)+\frac{m_D^2}{8 k\negcdot p}B_0(m_D^2,m_K^2,m_K^2)\\
&-\frac{p^2}{8 k\negcdot p}B_0(p^2,m_K^2,m_K^2)+\frac{1}{4}\Big]+ V_{ud} V_{cd}^*  \Big[ \quad m_K\to m_\pi, \;\Delta_s^*\to \Delta^* \quad \Big]\Big\},
\end{split}
\end{align}
\begin{align}
\begin{split}
C_{0kp}^{5\_1+5\_2}&= 2 i K {m_D}\Big\{ V_{us}V_{cs}^* \Big[\frac{m_K^2}{2}C_0(0,p^2,m_D^2,m_K^2,m_K^2,m_K^2)+\frac{p^2}{8 k\negcdot p}B_0(m_D^2,m_K^2,m_K^2)\\
&-\frac{p^2}{8 k\negcdot p}B_0(p^2,m_K^2,m_K^2)+\frac{1}{4}\Big]+ V_{ud} V_{cd}^*  \Big[ \quad m_K\to m_\pi, \;\Delta_s^*\to \Delta^* \quad \Big]\Big\},
\end{split}
\\
C_{0\eta}^{5\_3}&=-\frac{i K  m_D}{2} \Big[ V_{us}V_{cs}^* B_0(m_D^2,m_K^2,m_K^2)+  V_{ud}V_{cd}^* B_0(m_D^2,m_\pi^2,m_\pi^2)\Big],
\\
C_{0kp}^{5\_3}&=0,
\end{align}
where $\Delta_s^*=m_{D_s^*}-m_D$, $\Delta^*=m_{D^*}-m_D$, $K=\sqrt{m_D}{G_f}{a_1 e^3\alpha}/({16 \sqrt{2} \pi^2})$, while in $C_{\eta, kp}^{4\_4}$ we have used the  abbreviation
\begin{align}
\begin{split}
f_\eta(p,k,m,\Delta)=&\bar{C}_{00}(k,m,m,vp+\Delta)+(m^2-\Delta^2)\bar{D}_{00}(p,p+k,m,m,m,\Delta)\\
&-vp\; C_{001}(p^2,k^2,(p+k)^2,m^2,m^2,m^2)-m_D C_{002}(p^2,\dots)-\Delta C_{00}(p^2,\dots),
\end{split}\nonumber
\end{align}
\begin{align}
\begin{split}
f_{kp}(p,k,m,\Delta)=&m_D \bar{C}_{12}(k,m,m,vp+\Delta)+\bar{C}_{11}(k,m,m,vp+\Delta)\\
&+(m^2-\Delta^2)\Big[\bar{D}_{11}(p,p+k,m,m,m,\Delta)\\
&\qquad+m_D\big(\bar{D}_{12}(p,\dots)+2 \bar{D}_{13}(p,\dots)+\bar{D}_{1}(p,\dots)\big)\\
&\qquad+ m_D^2\big(\bar{D}_{23}(p,\dots)+\bar{D}_{33}(p,\dots)+\bar{D}_{3}(p,\dots)\big)\Big]\\
&-m_D^3\Big[\frac{1}{m_D^2} C_{001}(p^2,k^2,(p+k)^2,m^2,m^2,m^2)+\frac{2}{m_D^2}C_{002}(p^2,\dots)+C_{222}(p^2,\dots)\\
&\qquad\qquad+\frac{vp}{m_D}C_{112}(p^2,\dots)+\left(1+\frac{vp}{m_D}\right) C_{122}(p^2,\dots)+\frac{1}{m_D^2}C_{00}(p^2,\dots)\\
&\qquad\qquad +C_{22}(p^2,\dots)+\frac{vp}{m_D}C_{12}(p^2,\dots) \Big]\\
&-\Delta m_D^2\Big[ C_{22}(p^2,\dots)+C_{12}(p^2,\dots)+C_{2}(p^2,\dots)\Big],
\end{split}\nonumber
\end{align}
with the dots representing the same dependence on the arguments as for the first function in the square brackets.

\section{Nonresonant SD invariant amplitudes} \label{app-C}\setcounter{equation}{0}%
In this appendix we list the invariant amplitudes corresponding to the diagrams on Fig \ref{SDdiagr}. We use the notation of Eq. \eqref{inv_ampl}, where we write down only nonzero form factors
\begin{align}
C_0^{\text{SD}.1}&=i\frac{4}{3} K  \frac{V_{ub}V^*_{cb}}{v\negcdot k +\Delta^*}\left(\beta+\frac{1}{m_c}\right) \frac{(k\negcdot p)^2}{m_D^2}\frac{C_7-C_7'}{a_1},
\\
D_0^{\text{SD}.1}
&=\frac{4}{3} K \frac{V_{ub}V^*_{cb}}{v\negcdot k +\Delta^*}\Big(\beta+\frac{1}{m_c}\Big)\frac{v\negcdot p}{m_D}\frac{C_7+C_7'}{a_1},
\\
C_0^{\text{SD}.2}&= C_0^{\text{SD}.1} ( \rm{with}\; k \leftrightarrow p),
\\
D_0^{\text{SD}.2}&= D_0^{\text{SD}.1} ( \rm{with}\; k \leftrightarrow p),
\\
D_0^{\text{SD}.3}&= -\frac{1}{3}K \frac{ V_{ub}V_{cb}^*}{v\negcdot k+\Delta^*}\Big( \beta+\frac{1}{m_c}\big)\frac{p^2}{m_D}\frac{C_9+C_9'}{a_1} ,
\end{align}
\begin{align}
D_5^{\text{SD}.4}&= D_0^{\text{SD}.3} ( \rm{with}\; C_9^{(')} \to  C_{10}^{(')}),\\
M_{\text{BS}}^{\text{SD}.5a}+M_{\text{BS}}^{\text{SD}.5b}&=i \frac{1}{2} K V_{ub}V_{cb}^* \frac{m}{m_D}
\frac{C_{10}-C_{10}'}{a_1},
\end{align}
where $\Delta^*=m_{D^*}-m_D$ and  $K=\sqrt{m_D}{G_F}{a_1 e^3\alpha}/({16 \sqrt{2} \pi^2})$ have been used, while $m$ is the lepton mass. Note that the ``wrong chirality'' Wilson coefficients $C_{7,9,10}'$ are negligible in the SM.

\end{document}